\documentclass[preprint,superscriptaddress]{revtex4-2}

\usepackage{graphicx}
\usepackage{dcolumn}
\usepackage{bm}
\usepackage{hyperref}
\usepackage{indentfirst}
\usepackage[labelformat=simple]{subcaption}

\graphicspath{ {./figures/} }
\usepackage[section]{placeins}
\usepackage{xr-hyper}
\begin{document}
\title{Impact Dynamics of Droplet Containing Particle Suspensions on Deep Liquid Pool}

\author{Boqian Yan}
\affiliation{
 Department of Mechanical and Industrial Engineering, Northeastern University, Boston, USA, 02115.
}

\author{Xiaoyu Tang}
\email{x.tang@northeastern.edu}
\affiliation{
Department of Mechanical and Industrial Engineering,
Northeastern University, Boston, USA, 02115.
}
\affiliation{
Department of Chemical Engineering,
Northeastern University, Boston, USA, 02115.
}

\begin{abstract}

Droplets impacting on liquid pools have been extensively studied because of their complex dynamics and relevance to many natural and industrial processes. However, most of the research has focused on droplets of Newtonian fluids, and much less is known about droplets containing suspension particles, which adds another layer of complexity of non-Newtonian effect.
To address this gap, we studied droplets containing cornstarch particles impacting a deep water pool over a range of impact velocities $u_0$ ($0.5-3 \text{ m/s}$) and particle volume fractions $\phi$ ($36.8\% - 47.5\%$).
We identified five distinct impact phenomena, several of which are absent in the impacts of Newtonian droplets, and collected them in a regime map in the non-dimensional space of Weber number and $\phi$.
Through an energy-based scaling analysis, we elucidated the origins of and transitions between these phenomena, revealing that the competition between the pool cavity dynamics and the rheological properties of the suspension is the key factor governing the observed impact behaviors.
Our results deepen the understanding of the complex interplay between impact dynamics, cavity evolution, and suspension rheology, offering practical guidance for a range of engineering applications.


\end{abstract}


\maketitle

\section{\label{section1} Introduction}

The impact of droplets on a liquid pool plays a crucial role in many engineering and manufacturing processes \cite{yarin2006drop, rein1993phenomena, pan2007dynamics, che2018impact, kim2020raindrop, gart2015droplet, bagchi2021penetration, dressaire2016drop, pack2017failure, tsai2009drop, guemas2012drop, jain2019deep}. The interplay of inertial, viscous, and capillary forces produces a range of outcomes, such as bouncing \cite{tang2019bouncing, richard2002contact, de2015wettability, kolinski2014drops, tang2018bouncing, zou2011experimental, he2021drop}, spreading \cite{lee2016universal, lagubeau2012spreading, roisman2002normal, laan2014maximum, wildeman2016spreading, tang2019spreading}, and splashing \cite{mani2010events, xu2005drop, zhang2010wavelength, murphy2015splash, yang2021experimental, juarez2012splash, peng2021droplet}.
By controlling these dynamics, engineers can manipulate the mass transfer between the droplet and the pool to improve both the yield and the product quality.
However, in many applications, such as material fabrication \cite{liu2011extension}, droplets contain particles for additional functionality. The suspension droplets often exhibit non-Newtonian behaviors \cite{mewis2012colloidal, morris2020toward, bi2011jamming}, such as viscoelasticity, shear thinning, and shear thickening, adding rheological complexity to the impact dynamics.
On the other hand, the impact process generates complex flow fields inside the droplet and within the pool, which redistribute the particles and create heterogeneous viscosity distributions.
Therefore, a thorough investigation of suspension droplets impact dynamics not only guides practical applications but also advances our fundamental knowledge of suspension behaviors in complex flow fields, shedding light on phenomena such as impact-induced jamming \cite{waitukaitis2012impact, han2016high, han2019dynamic}.

The impact dynamics of suspension droplets on solid substrates have been studied both experimentally \cite{shah2024drop, Boyer2016Drop, bertola2015impact} and numerically \cite{cao2024regulating}, revealing unique phenomena compared to Newtonian droplet impact. Boyer et al. \cite{Boyer2016Drop} showed that the final deformation of the droplet is insensitive to the initial impact velocity due to the suspension rheology with Bagnoldian relationship.
Shah et al. \cite{shah2022coexistence, shah2024drop} demonstrated that, at high particle volume fractions, droplets behave like elastic solid spheres and the jamming occurs within milliseconds after the initial contact.
They also observed the coexistence of liquid and solid phases at intermediate volume fractions.
These studies on the impacts of suspension droplets on a solid substrate unveiled the rich physics stemming from the coupling between suspension rheology and droplet impact. Moreover, the impact on a liquid pool adds another layer of complexity and remains less explored. A liquid surface can deform upon droplet impact which may provide less stress to the suspension, while the droplet can penetrate the pool and release the particles to the liquid to interact with the resulting flow field. These unique features raise key questions: What impact phenomena would emerge when suspension droplets impact on liquid pools, and how do their rheological properties influence such behaviors?

To address these questions, we investigated the impact dynamics of cornstarch suspension droplets on a deep water pool using high-speed imaging and rheological characterization. 
By varying the impact velocity $u_0$ and particle volume fraction $\phi$ of the droplets,
we identified impact phenomena that are unique in the suspension droplets impact on liquid pool such as \textit{Wrapped Bubble} and \textit{Solid Lump}, and organized those in a regime map of Weber number and the particle volume fraction. 
We examined and unveiled how non-Newtonian behavior competes with impact inertia and capillarity in setting the observed outcomes. 
By combining rheological measurements and energy balance analysis, we identified theoretical predictions for the transition boundaries between different regimes, thus providing guidelines for selecting conditions to obtain desired impact outcomes for practical applications. 
Overall, this work provides a framework for interpreting and predicting the impact dynamics of suspension droplets on deformable liquid surfaces, and highlights the role of suspension rheology in shaping droplet–pool coupling and regime transitions.

\section{\label{section2} Methods }

\subsection{\label{section2A} Materials and Characterization}
The suspension fluid is composed of cornstarch particles (Sigma Aldrich) with density ${{\rho }_{c}}=1630\text{ kg}/{{\text{m}}^{\text{3}}}$, and a solution of 55 wt.$\%$ Cesium Chloride (CsCl, Sigma Aldrich) in DI water. The density of the suspending liquid matches that of the particle such that the suspension remains stable and uniformly distributed for up to a week. Cornstarch particles are relatively monodisperse with a characteristic diameter of $14 {\text{ $\mu$} \text{m}}$  \cite{Fall2012Shear} and Brownian motion can be neglected \cite{Tanner2018Review}.
In this work, we report the concentration of cornstarch particles in terms of the volume fraction $\phi$ (Table \ref{tab:20}) rather than the mass fraction because the rheological properties of suspensions are primarily reported by their particle volume fractions. The conversion from mass fraction to volume fraction follows Han et al.\cite{Han2017Measuring}, which is detailed in Sec. S1 of Supplementary Material \cite{supplemental}. The volume fraction of the cornstarch particles used in the droplet impact experiments ranges from 36.8$\%$ to 47.5$\%$, which are below the jamming threshold $\phi_{J}=52.0\%$, tested during sample preparation.

\begin{table}[h]
    \caption{\label{tab:20}
    Range of weight percentage and volume fraction of the cornstarch suspensions tested in this work.}
    \begin{ruledtabular}
    \begin{tabular}{c|cccccccccc}

    \textbf{ Weight percentage $wt. $\%$ $} & 30.0 & 31.0 & 32.0 & 33.0 & 34.0 & 35.0 & 36.0 & 37.0 & 38.0 & 39.0 \\
    \hline
    \textbf{Volume fraction $\phi$ [$\%$]} & 36.8 & 38.0 & 39.2 & 40.4 & 41.6 & 42.8 & 44.0 & 45.2 & 46.3 & 47.5 \\
    \end{tabular}
    \end{ruledtabular}
\end{table}
\FloatBarrier

Rheological properties of suspensions at various volume fractions were measured using a DHR-3 rheometer (TA Instruments) in a parallel-plate configuration (40 mm plate diameter, $500\,\mu\mathrm{m}$ gap). A logarithmically spaced shear-rate sweep was applied from \(1 \, \text{s}^{-1}\) to \(5000 \, \text{s}^{-1}\) and  the corresponding shear stress was recorded. The results and interpretations will be discussed in Sec. \ref{section3} B.

\begin{figure}[htb]
	\centering
	\includegraphics[width=0.6\linewidth]{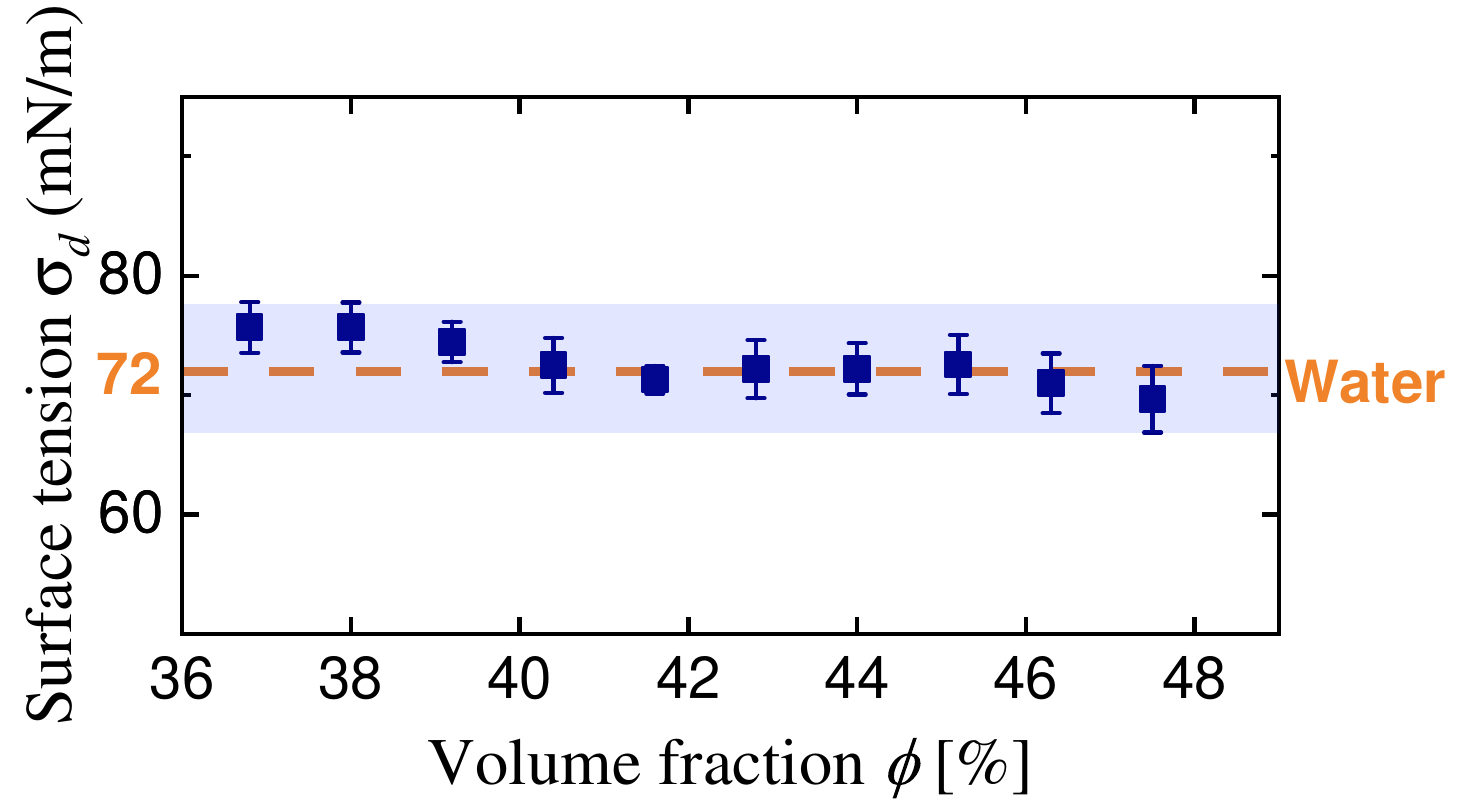}
    \caption{Surface tension measurement of the cornstarch suspension $\sigma_d$ with various volume fractions $\phi$ by the pendant drop method. The blue shaded area highlights the range of $\sigma_d$ with all $\phi$. The orange dash line marks the value of surface tension of DI water.} \label{fig_surface_tension}
\end{figure}

The surface tension of the cornstarch suspension $\sigma_d$ was measured using the pendant drop method \cite{Stauffer1965The, arashiro1999use} at different volume fractions $\phi$ and is shown in Fig. \ref{fig_surface_tension}. There is no significant change in ${\sigma}_{d}$ with varying $\phi$, and the values hover around the value of water ${\sigma}_{w}$, indicated by the orange dashed line. Thus, both ${\sigma}_{d}$ and ${\sigma}_{w}$ are denoted as $\sigma$, with the value ${{\sigma}}=72\text{ mN/m}$ for subsequent analysis.

\subsection{\label{section2B} Experimental Set-up and Procedure}
\begin{figure}[htb]
	\centering
		\centering
		\includegraphics[width=0.7\textwidth]{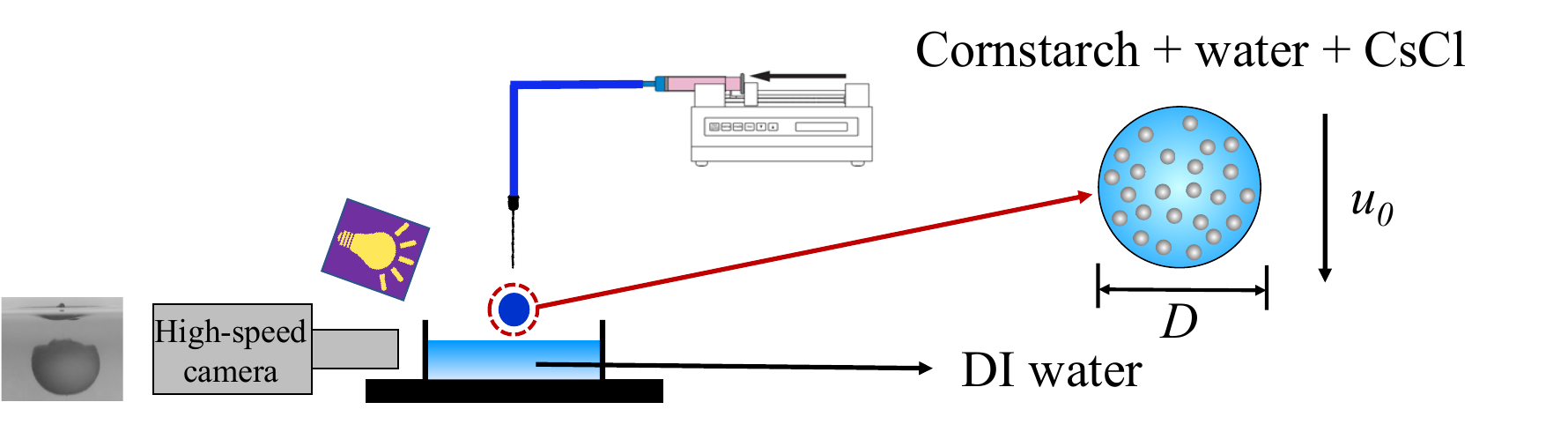}
    \caption{Experimental configuration for studying cornstarch suspension droplets impact on a water pool.} \label{fig_setup}
\end{figure}

To prepare the suspension fluid, cornstarch particles were mixed with the CsCl solution using a glass stirrer and the mixture was sonicated (Branson 8510) for 10 minutes. The suspension was then transferred to a syringe connected to a needle (Gauge 18, McMaster-Carr). To ensure consistent results, each batch of suspension fluid was used within three hours of preparation. As shown in Fig. \ref{fig_setup}, the syringe was mounted on a syringe pump (New Era), which delivered the fluid at a low constant flow rate. Droplets were detached from the needle  once their weight exceeded the capillary force at the tip of the needle and fell on a water pool in a cubic container (50$\times$50$\times$50 $\mathrm{ mm^3}$). The suspension droplets remain spherical prior to the impact, and their diameter $D$ ranges from 2.8 mm to 3.2 mm for all volume fractions, as shown in Fig.S1(a) of Supplemental Material~\cite{supplemental}.
The large pool height $H = 50\, \mathrm{ mm}$  ($H/D > 10$) ensures that the container substrate does not affect impact dynamics \cite{tang2018bouncing}.
The impact velocity was controlled by varying the needle height above the pool from 3 mm to 400 mm, which gives a velocity ranging from 0.5 m/s to 3 m/s. A high-speed camera (Phantom VEO 1310, Vision Research) recorded the side-view images of each impact at 10000 fps with front lighting.

\FloatBarrier

\subsection{\label{section2C} Image Processing}

The images were post-processed with customized MATLAB codes to extract key parameters such as the droplet diameter $D$ and the impact velocity $u_0$. $u_0$ is obtained by extracting the slope of a linear fit of the center positions $y$ of the droplet versus time $t$, as shown in Fig.S1(b) of Supplemental Material~\cite{supplemental}, which confirms that the droplets fall at a constant velocity $u_0$ prior to impact.

The positions of the top and bottom points of the droplet are also tracked from the moment of first contact until full immersion in the pool to monitor the deformation as the droplet penetrates through the liquid surface, as shown in Fig.S2 of Supplemental Material~\cite{supplemental}.
The linear behavior at each stage for each line confirms the constant velocities of the top and bottom of the droplet during penetration. By linear fitting of each position versus time, the velocities of the top $u_{top}$ and bottom $u_{bot}$ are extracted.

\section{\label{section3} Results}

\subsection{\label{section3B} Impact Phenomenology}

\begin{figure}[htb]
	\centering
	\begin{subfigure}{1.0\linewidth}
		\centering
		\includegraphics[width=0.77\linewidth]{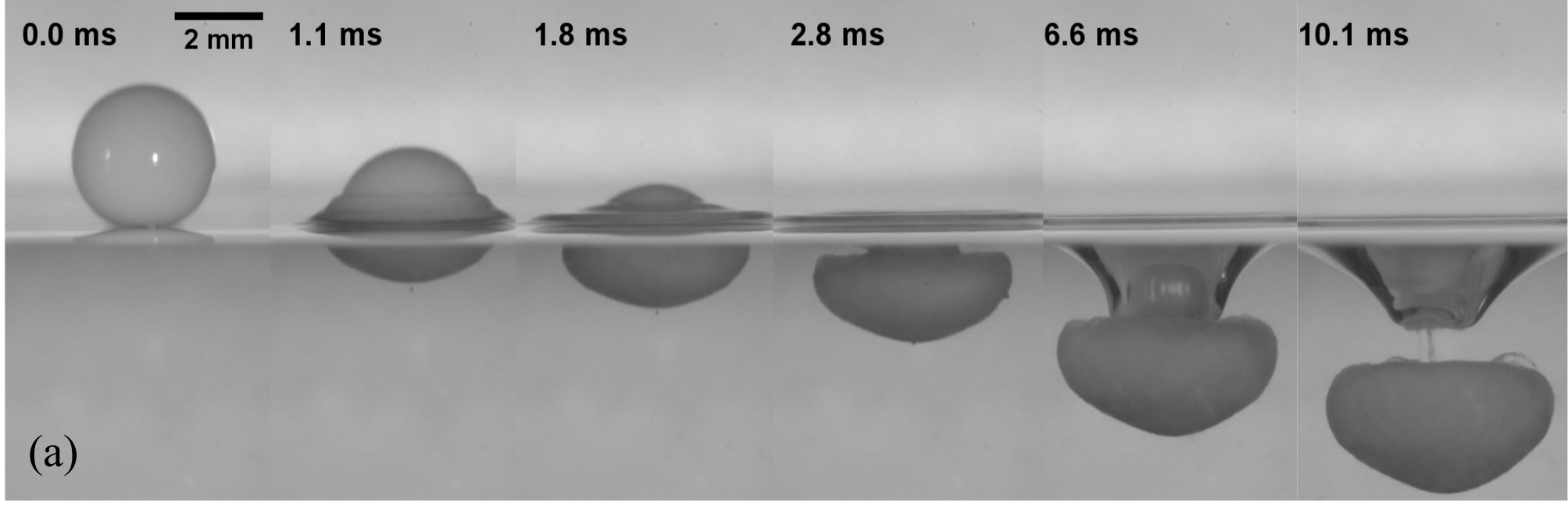}
	\end{subfigure}
	\begin{subfigure}{1.0\linewidth}
		\centering
		\includegraphics[width=0.77\linewidth]{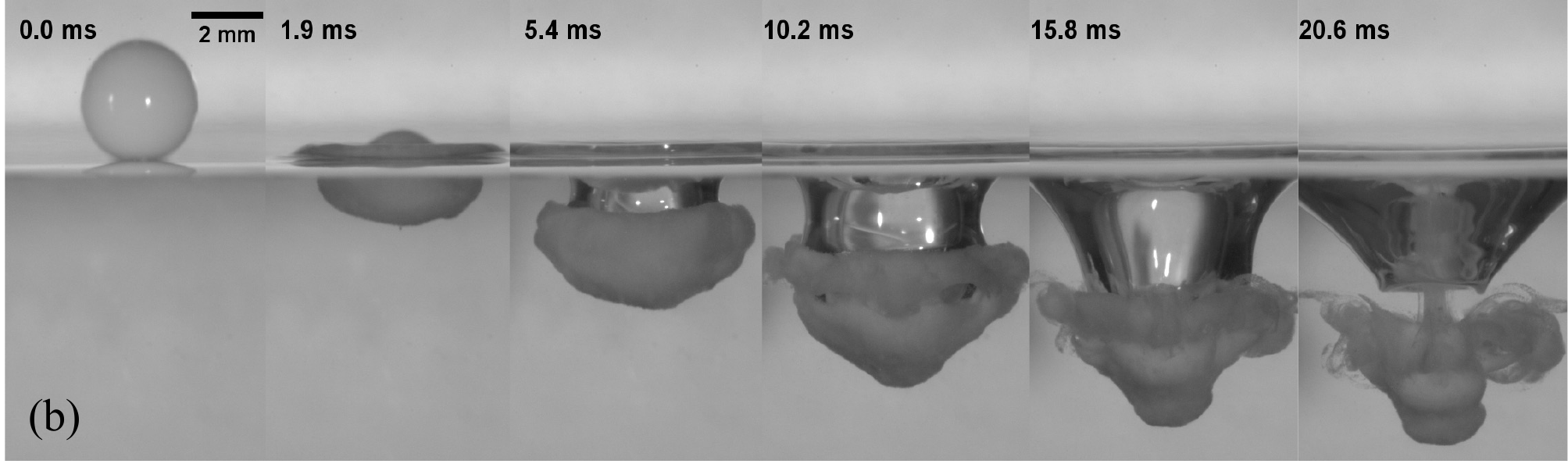}
	\end{subfigure}
    \begin{subfigure}{1.0\linewidth}
		\centering
		\includegraphics[width=0.77\linewidth]{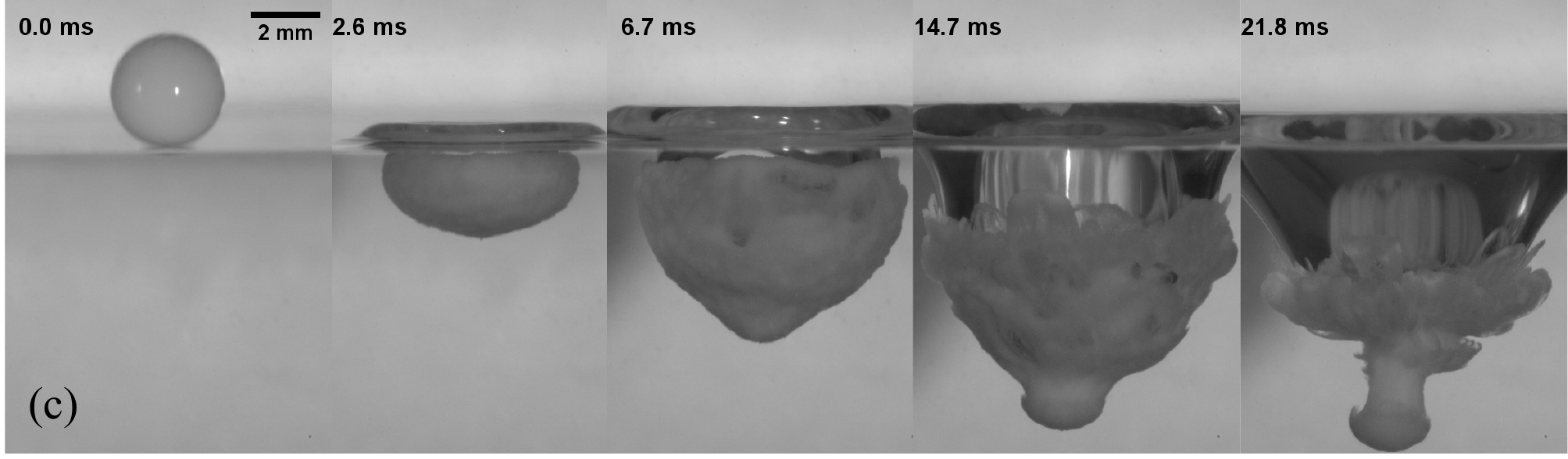}
	\end{subfigure}
    \begin{subfigure}{1.0\linewidth}
		\centering
		\includegraphics[width=0.77\linewidth]{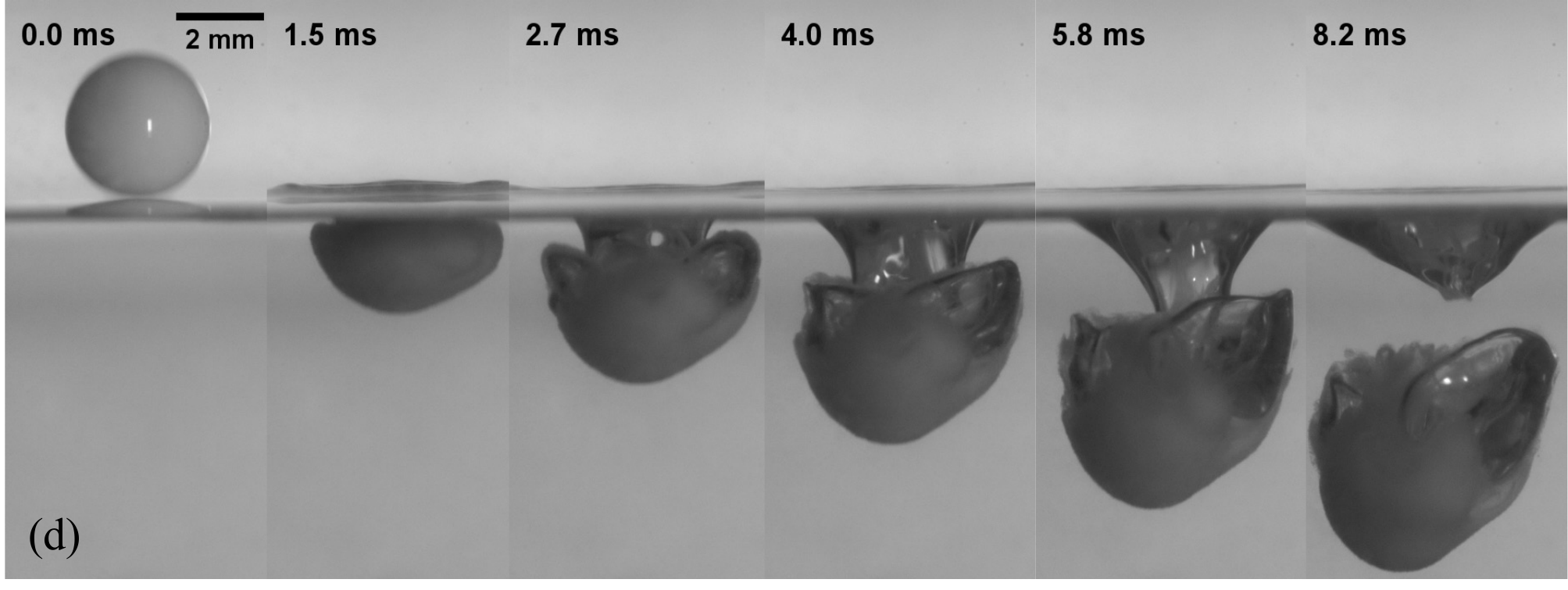}
	\end{subfigure}
    \begin{subfigure}{1.0\linewidth}
		\centering
		\includegraphics[width=0.77\linewidth]{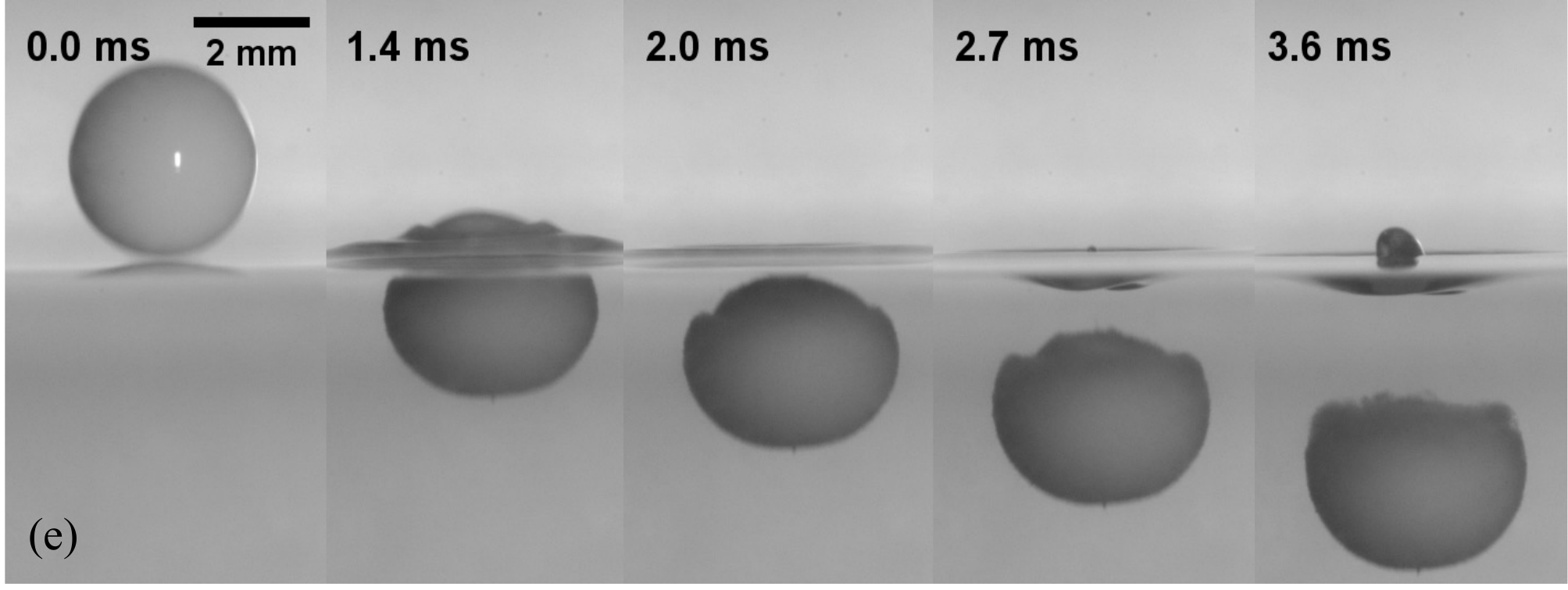}
	\end{subfigure}
    \caption{Snapshots of the impact dynamics of cornstarch droplets on a deep water pool. (a) \textit{Liquid Lump} ($\phi =36.8 $\%$ $, $u_0=0.85 \,\mathrm{m/s}$); (b) \textit{Partially Attached} ($\phi =36.8 $\%$ $, $u_0=1.41 \,\mathrm{m/s}$); (c) \textit{Fully Attached} ($\phi =36.8 $\%$ $, $u_0=2.07 \,\mathrm{m/s}$); (d) \textit{Wrapped Bubble} ($\phi =41.6 $\%$ $, $u_0=1.45 \,\mathrm{m/s}$); (e) \textit{Solid Lump} ($\phi =47.5 $\%$ $, $u_0=1.39 \,\mathrm{m/s}$)} \label{fig_observation}
\end{figure}

Fig. \ref{fig_observation} shows five distinct impact dynamics of cornstarch suspension droplets on a water pool at various  $u_0$ and volume fractions $\phi$. At low $u_0$ (0.85 m/s) and $\phi$ ($36.8\%$) as shown in Fig. \ref{fig_observation}(a), the suspension droplet merges with the liquid pool upon impact, detaches from the cavity, and the mass (solution and cornstarch particles) from the droplet continues to descend due to inertia. Once the top surface of the droplet moves below the original pool surface, the cavity forms and starts to grow and interact with the mass from the droplet. While the cavity retracts due to the capillary effect, the mass from the droplet continues to descend, forming a liquid lump that can be distinguished from the liquid pool. This regime is named \textit{Liquid Lump}.
At higher $u_0$ (1.41 m/s) with the same $\phi$ as shown in Fig. \ref{fig_observation}(b), more kinetic energy in the droplet converts to more energy in the cavity, leading to a stronger interaction with the mass from the droplet. The mass from the droplet is partially dispersed and wrapped around part of the cavity, but the bottom portion of the droplet remains together in a hemispherical shape,
as the cavity grows downward in a cylindrical fashion. This regime is named \textit{Partially Attached}, emphasizing the intermediate effect of the interaction between the cavity and the mass from the droplet.
At even higher $u_0$ (2.07 m/s) as shown in Fig. \ref{fig_observation}(c), the cavity gains so much more energy from the suspension droplet that it expands spherically dominated by the inertia effect. The mass from the droplet is fully dispersed into a hemispherical shell and wraps around the entire cavity.
This regime is named \textit{Fully Attached}.
The quantitative distinction between \textit{Partially Attached} and \textit{Fully Attached} regimes is determined from the experimental measurement of the evolution of the cavity shape, with the details provided in Supplemental Material~\cite{supplemental}, Sec.~S3.
It is noted that the cavity behaviors in these three regimes are similar to those of Newtonian droplets: with increasing impact speed, the cavity size grows, and its shape transitions from cone to hemisphere. Snapshots for water droplet impact are shown in Fig.S4 of Supplemental Material~\cite{supplemental} for qualitative comparison.

Distinctive phenomena emerge at higher volume fraction $\phi$. At similar $u_0$ (1.45 m/s) as in the \textit{Partially Attached} case ($\phi=36.8\%$), \textit{Wrapped Bubble} occurs at $\phi =41.6 \%$ as shown in Fig. \ref{fig_observation}(d) and \textit{Solid Lump} is observed at $\phi=47.5 \%$ as shown in Fig. \ref{fig_observation}(e). For the \textit{Wrapped Bubble} case, the cavity becomes unstable shortly after the impact, in contrast to the continuous expansion in previous cases. The lower section of the mass from the droplet remains undisturbed, descending as a cohesive lump; while the upper section of the mass from the droplet wraps around the cavity, eventually pulls one or more bubbles from the cavity and brings them down together with the droplet mass. The bubble formation mechanism differs fundamentally from that in Newtonian droplet impacts, as we will discuss in Sec. \ref{section4B}. For the \textit{Solid Lump} case, the entire mass from the droplet descends as a whole, maintaining an almost unchanged shape. In addition, minimal cavity formation is observed throughout the impact process, which resembles the impact of hydrophilic solid balls on liquid surfaces \cite{duez2007making, truscott2014water, aristoff2010water, marschall2003cavitation}.

\FloatBarrier

\begin{figure}[htb]
	\centering
	\includegraphics[width=0.5\linewidth]{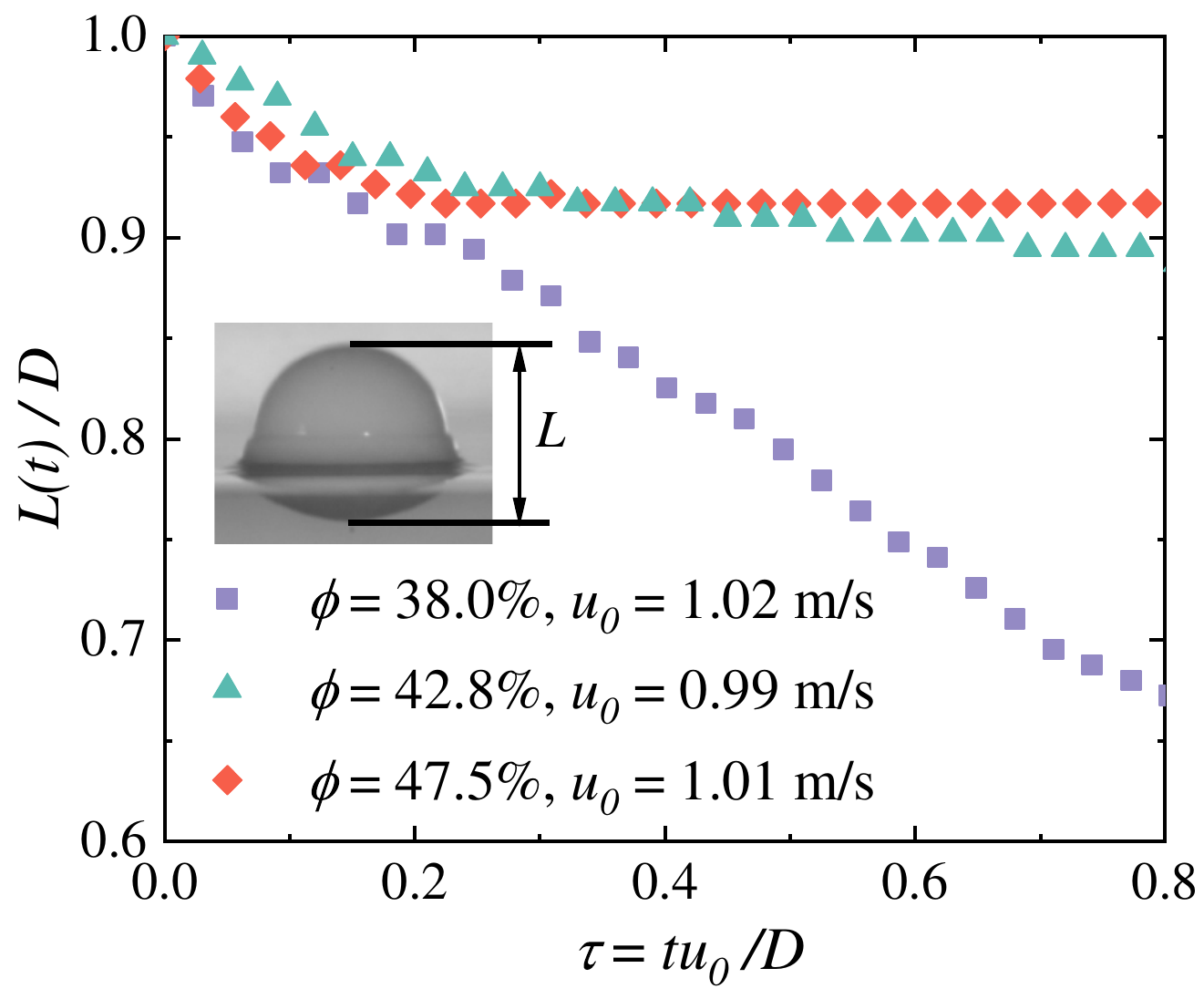}
    \caption{Variations of the normalized height of droplet $L/D$ with normalized time $\tau =t{{u}_{0}}/D$ for low $\phi$ ($=38.0\%$), intermediate $\phi$ ($=42.8\%$) and high $\phi$ ($=47.5\%$) at ${{u}_{0}} \approx 1.0\text{ m/s}$.} \label{fig_deformation}
\end{figure}

In both high $\phi$ cases, the mass from the droplet exhibits jamming behavior. The jamming transition has been observed for the impact of suspension droplets on solid surfaces, manifested as limited deformation \cite{shah2022coexistence, Boyer2016Drop}. To validate the jamming transition at high volume fractions, the vertical length of the droplets, $L(t)$, is measured from the moment of first contact with the pool surface to full immersion. Fig. \ref{fig_deformation} presents the normalized vertical length $L/D$ as a function of the normalized time $\tau =t{{u}_{0}}/D$. At $\phi = 38.0 \%$,  $L/D$ decreases monotonically, indicating a continuous deformation similar to a Newtonian droplet. At $\phi =42.8 $\%$ $, the deformation exhibits a step-wise behavior.
This step-wise behavior is unique to suspension droplets impacting liquid pools and has not been reported for impacts on solid substrates. The difference may arise because the solid substrate imposes a rigid geometric constraint, whereas a liquid pool provides a compliant resistance without geometric constraint.
At $\phi =47.5 $\%$ $, $L/D$ reaches a plateau shortly after contact, indicating that the deformation ends due to jamming. This observation is consistent with the abrupt cessation of deformation for suspension droplets impacting solid substrates \cite{Boyer2016Drop, shah2022coexistence}.

\FloatBarrier

\begin{figure}[htb]
	\centering
	\includegraphics[width=0.7\linewidth]{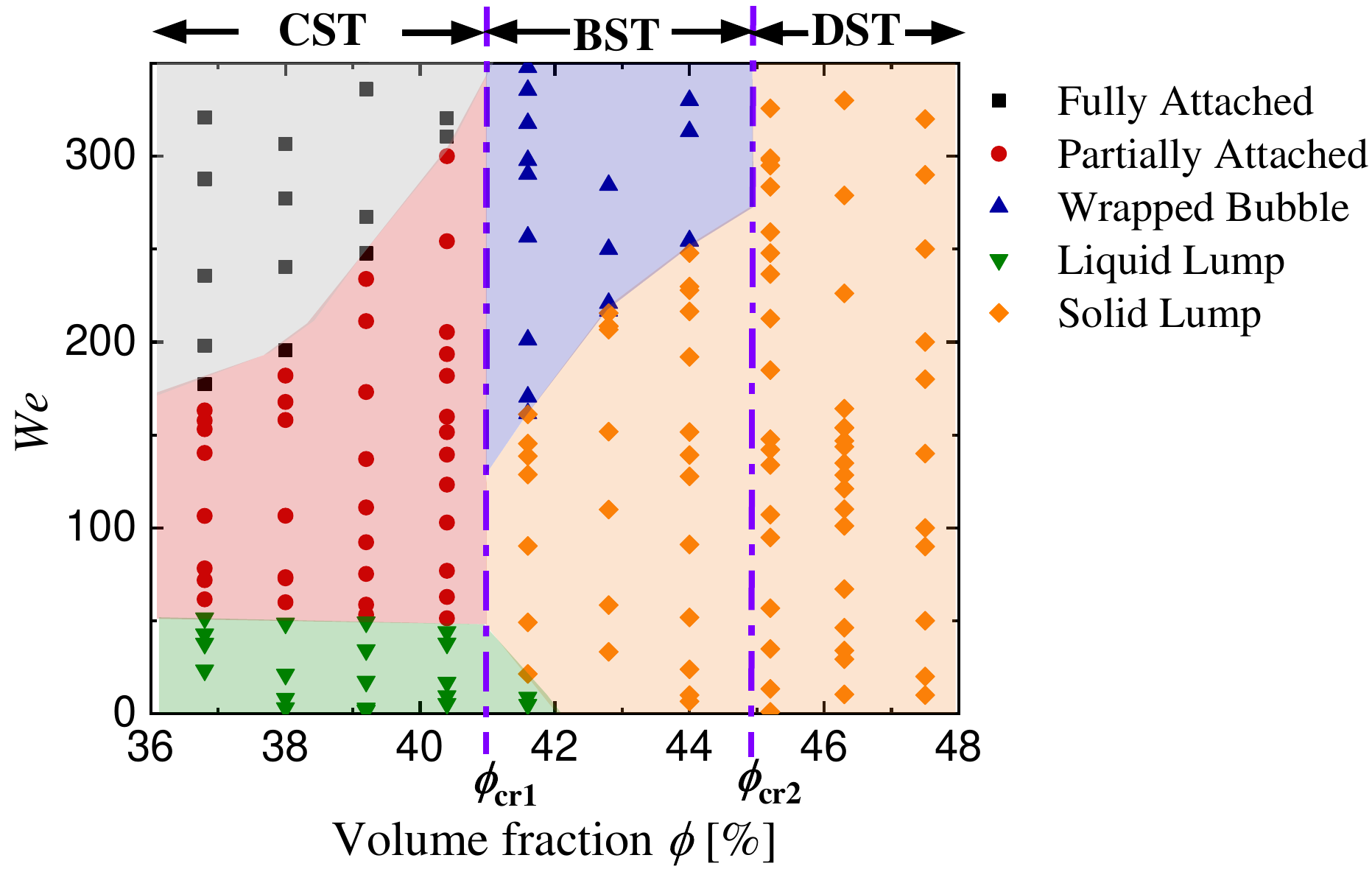}
    \caption{Regime map of impact outcomes of cornstarch suspension droplets on a deep water pool in the $We$-$\phi$ space. Two dash-dotted lines denote two critical values of volume fractions $\phi_{cr1} \approx 41\% $ and $\phi_{cr2} \approx 45\%$ that divide the regime map into three regions based on experimental observations.} \label{fig_regime}
\end{figure}

The five impact phenomena are organized into a regime map in the space of particle volume fraction $\phi$ and Weber number $We = \rho_du_0^2D/\sigma$ (the ratio between the inertia and the capillarity of the droplet), as shown in Fig.\ref{fig_regime}.
$We$ is selected as the dimensionless parameter to capture the dominance of capillarity of the pool surface at low $u_0$ and the dominance of inertia at high $u_0$.
Two critical volume fractions, $\phi_{cr1} \approx 41\%$ and $\phi_{cr2}\approx45\%$, can be identified that divide the map into three regions, as shown by the dash-dotted lines.
At low volume fraction ($\phi <\phi_{cr1}$), increasing $We$ drives the transition from \textit{Liquid Lump} to \textit{Partially Attached} to \textit{Fully Attached}. Both \textit{Wrapped Bubble} and \textit{Solid Lump} are observed in the intermediate $\phi$ range ($\phi_{cr1}<\phi <\phi_{cr2}$), while only \textit{Solid Lump} is observed for $\phi>\phi_{cr2}$. Transition mechanisms will be identified to predict the transition boundaries in the following sections.

\FloatBarrier

\subsection{\label{section3C} Rheological Properties}

\begin{figure}[htb]
	\centering
	\includegraphics[width=0.85\linewidth]{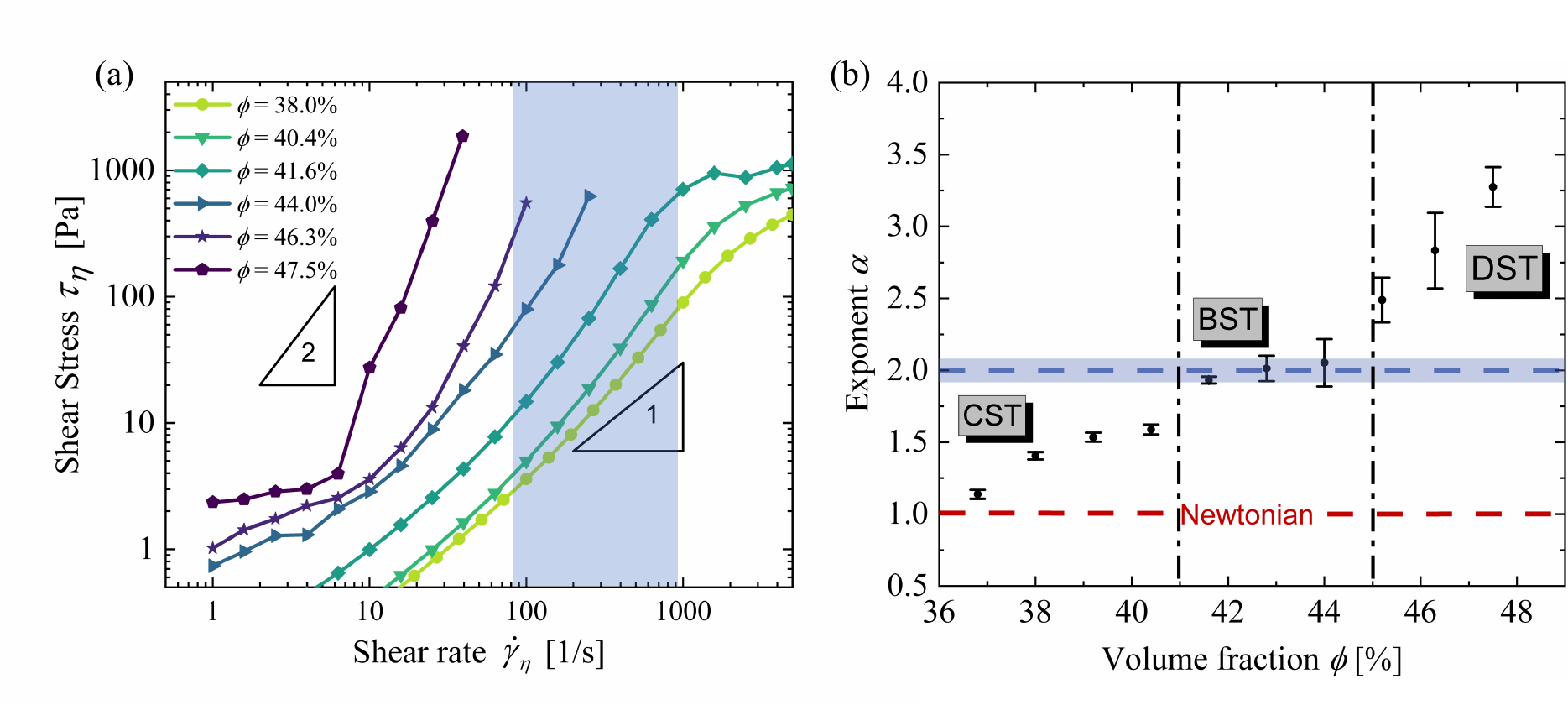}
    \caption{(a) Rheological measurement of cornstarch suspensions with various volume fractions $\phi$. (b) Variation of exponent $\alpha $ in rheological scaling law $\tau_{\eta} \propto {{\dot{\gamma }_\eta}^{\alpha }}$ with various volume fractions $\phi$.
    The error bars represent the fitting uncertainty of $\alpha$, obtained from the standard error of the power law fit of $\tau_\eta$ versus $\dot{\gamma}_\eta$ within the selected fitting range.} \label{fig_rheolgoy}
\end{figure}

The critical volume fractions are identified by characterizing the rheology of cornstarch suspensions in the range of $\phi$ tested in droplet impact experiments, as shown in Fig.~\ref{fig_rheolgoy}(a). As $\phi$ increases, the shear thickening behavior becomes more prominent. For high volume fractions, no data was obtained at high shear rate due to jamming, which leads to patches of local solidification and slipperiness, resulting in the termination of the measurements\cite{shah2022coexistence}.
The characteristic shear rate during the droplet impact process is estimated by $\dot{\gamma}={{u}_{0}}/D$ \cite{Boyer2016Drop, shah2022coexistence} and ranges from 80 to 900 $\mathrm{s^{-1}}$, which is marked by the shaded region in Fig. \ref{fig_rheolgoy}(a).
The data in the shaded region is fitted to the power law $\tau_{\eta} \propto {{\dot{\gamma }}_{\eta}^{\alpha }}$ \cite{Brown2014Shear}, where the exponent $\alpha$ is extracted and plotted against $\phi$ in Fig. \ref{fig_rheolgoy}(b). Since there is limited data in the shaded area for $\phi \geq 46.3\% $, the values of $\alpha$ are calculated based on the data outside the shaded region showing obvious shear thickening behavior.
The exact values of $\alpha$ for these cases do not affect the categorization, as discussed below.
The rheological behaviors of cornstarch suspensions are categorized based on Brown et al. \cite{Brown2014Shear, morris2020shear}. $\alpha < 2$ corresponds to Continuous Shear Thickening (CST), where the viscosity of the suspension increases gradually as the shear rate increases. Discontinuous Shear Thickening (DST) occurs for $\alpha > 2$, where the viscosity of the suspension increases rapidly.
Cases with $\alpha = 2$ are termed Bagnoldian Shear Thickening (BST), as they satisfy the Bagnoldian relationship \cite{bagnold1954experiments}.
Note that the term BST is used here to denote the $\alpha = 2$ rheology, while the indication of particle inertia is not relevant and thus not considered in the analysis of droplet impact dynamics.
Based on these criteria and $\alpha$ values extracted from the shear rate range during droplet impact, cornstarch suspensions with $\phi < 41\%$ and $\phi > 45\%$ present CST and DST behaviors, respectively, while cornstarch suspensions with $41\% < \phi < 45\%$ are categorized as BST since $\alpha$ remains close to 2, which is consistent with the work by Boyer et al.~\cite{Boyer2016Drop}. The categorization is summarized in Table \ref{tab:1} and shown in Fig. \ref{fig_rheolgoy}(b)
separated by the dash-dotted lines.
The error bars in Fig.~\ref{fig_rheolgoy}(b) represent the fitting uncertainty in $\alpha$ and show that the qualitative categorization is not affected by this uncertainty.
Note that the divisions between CST, BST, and DST regions should not be interpreted as sharply defined phase boundaries or step-like transitions, but rather as rheological categories
to highlight the evolution of rheological responses with $\phi$ and its correlation with the observed changes in the droplet impact regimes.
Thus,
two critical volume fractions: $\phi_{cr1} \approx 41$\%$ $ and $\phi_{cr2} \approx 45$\%$ $, can be identified to group the rheological behaviors of cornstarch suspensions into CST, BST, and DST regions.
Remarkably, these critical volume fractions based on rheology coincide exactly with those identified from the regime map of droplet impact behaviors in Fig.\ref{fig_regime}. This alignment clearly shows that critical volume fractions are rheology controlled. The regimes of \textit{Liquid Lump}, \textit{Partially Attached}, and \textit{Fully Attached} sit in the CST region, in which the impact behavior mostly resembles that of Newtonian droplet. In the DST region, only \textit{Solid Lump} occurs due to impact induced jamming. In the BST region, both \textit{Wrapped Bubble} and \textit{Solid Lump} occur, which exhibit the transitional behavior between CST and DST. The $We$ dependence in each rheological region will be discussed in the following Discussion section.

\FloatBarrier

\begin{table}[h]
    \caption{\label{tab:1}
    Rheological categorization of the cornstarch suspensions.
    }
    \begin{ruledtabular}
    \begin{tabular}{ccc}
    \textbf{Range of volume fraction $\phi$} & \textbf{Values of $\alpha$} & \textbf{Properties of fluid} \\
    \hline
    $\phi < 41$\%$ $ & $1<\alpha <2$ &	Continuous Shear Thickening (CST) \\
    $41$\%$ < \phi < 45$\%$ $ & $\alpha =2$	& Bagnoldian Shear Thickening (BST) \\
    $\phi > 45$\%$ $ & $\alpha >2$	& Discontinuous Shear Thickening (DST)\\
    \end{tabular}
    \end{ruledtabular}
\end{table}
\FloatBarrier

\section{\label{section4} Discussions}

\subsection{\label{section4A} Continuous Shear Thickening (CST) region}

In the CST region, the suspension exhibits slight shear thickening behavior, which does not significantly affect the interaction between the suspension droplets and the liquid pool. Thus, the observed regimes of \textit{Liquid Lump}, \textit{Partially Attached}, and \textit{Fully Attached} resemble the phenomenon of a Newtonian droplet impact, with the snapshots shown in Fig.S4 of Supplemental Material~\cite{supplemental}. To further quantify the similarity, the velocity of the bottom point of the mass from the droplet, $u_{bot}$, is measured and plotted against the impact velocity $u_0$, as shown in Fig.\ref{fig_ubot}(a). The slope gives $\beta = u_{bot} / u_0 = 0.61 \pm 0.009$, which is in good agreement with the theoretical value of $\beta_{N,theory} = 0.59$ derived for Newtonian droplets impacting on liquid pools \cite{fudge2021dipping}.

\begin{figure}[htb]
	\centering
	\includegraphics[width=0.85\linewidth]{figures/figure_ubot.png}
    \caption{ (a) Variations of $u_{bot}$ with respect to $u_0$ in CST region. (b) Variations of $u_{bot}$ with respect to $u_0$ in BST and DST regions.}
    \label{fig_ubot}
\end{figure}
\FloatBarrier

\begin{figure}[htb]
	\centering
	\begin{subfigure}{0.85\linewidth}
		\centering
		\includegraphics[width = 1.0\linewidth]{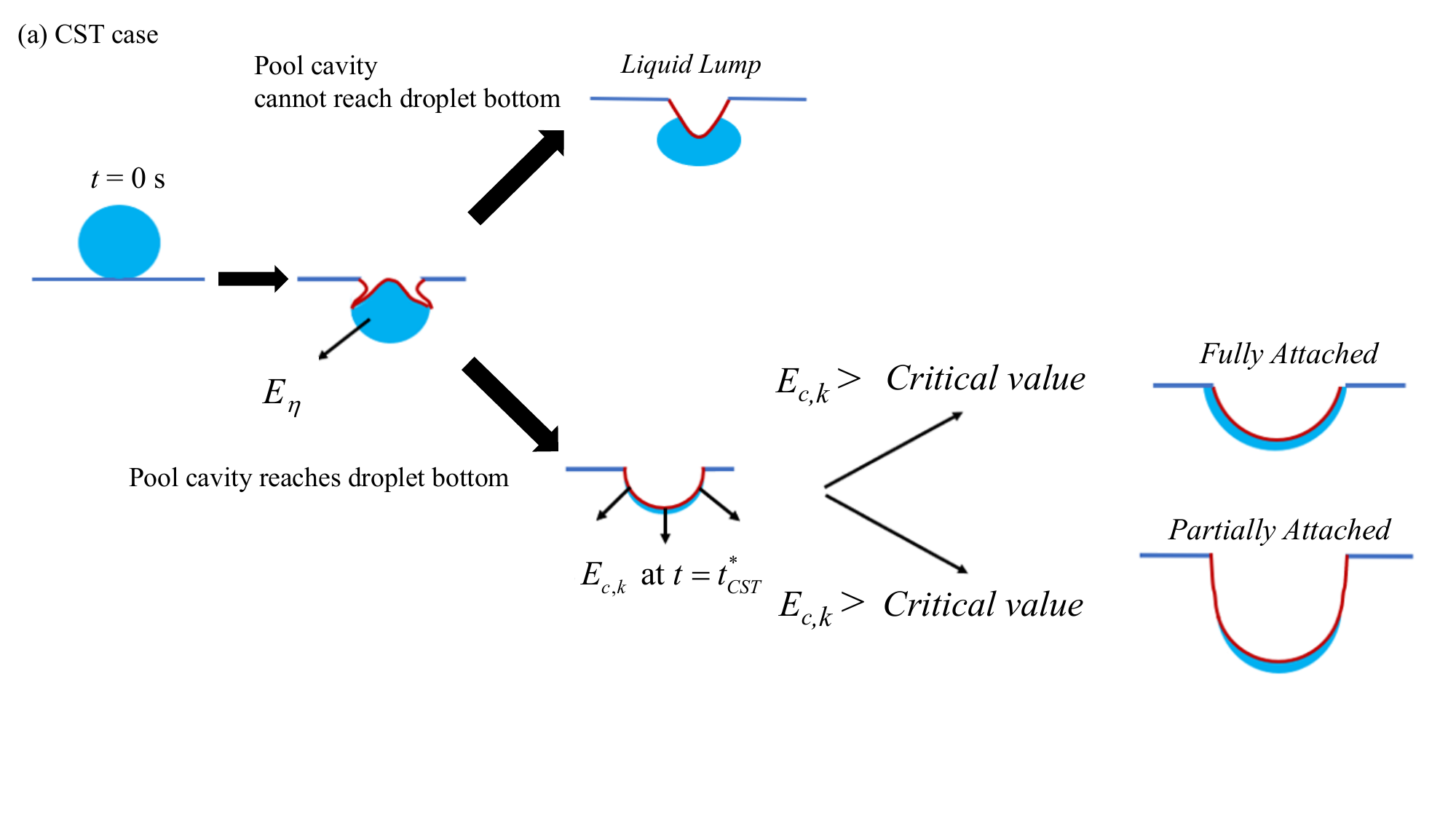}
	\end{subfigure}
	\begin{subfigure}{0.85\linewidth}
		\centering
		\includegraphics[width = 1.0\linewidth]{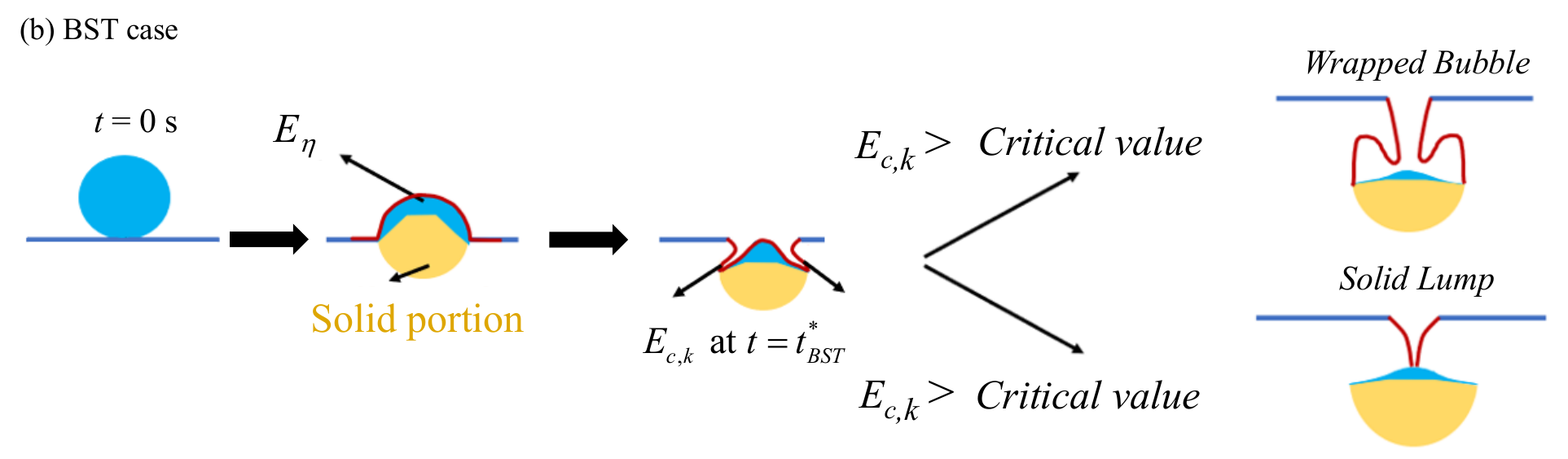}
	\end{subfigure}
    \caption{
    (a) Impact process within the CST region: The transition criterion is controlled by the cavity kinetic energy $E_{c,k}$ at $t_{CST}^*$, the moment when the cavity catches up with the bottom of the droplet. (b) Impact process within the BST region: the transition criterion is controlled by  $E_{c,k}$ at $t_{BST}^*$, the moment when the cavity aligns with the undisturbed pool surface where the cavity is nascent. Blue area: Liquid portion. Yellow area: Solid portion. Red line: cavity surface, which is the top surface of the droplet. Blue line: pool surface.
    }
    \label{TransitScheme}
\end{figure}

The transition between different regimes based on the droplet behaviors stems from the interaction between the cavity and the droplet. This interaction is best observed for water droplet impact, as shown in Fig.S4 of Supplemental Material~\cite{supplemental}, where the droplet is transparent in contrast to the opaque suspension droplet, while dyed to track the mass from the droplet. The observed interactions between the cavity and the Newtonian droplet will guide the analysis of the impact phenomenon of the suspension droplets in the CST region due to the similarity established above.
Once a water droplet impacts on the surface at $t=0$ s, it merges with the pool surface and the mass from the droplet descends through the pool. The cavity is formed by the top surface of the droplet as it expands downward.
At low impact speed, capillarity controls the cavity expansion; thus the cavity exhibits a cone shape (Fig.S4(a) of Supplemental Material~\cite{supplemental}).
At high impact speed, inertia dominates and the cavity forms a spherical cap undergoing radial expansion (Fig.S4(c) of Supplemental Material~\cite{supplemental}). As both the cavity and the mass from the droplet move downward, the cavity can catch up with the bottom of the mass from the droplet, which strongly disperses the mass into a thin film around the cavity as it expands.
Fig.\ref{TransitScheme}(a) illustrates the origin of each suspension droplet impact regime and the transition criteria between them in the CST region.
If the cavity does not have enough energy to catch up with the bottom of the mass from the droplet, the droplet remains intact, falling as a cohesive mass in the \textit{Liquid Lump} regime. The transition to the \textit{Partially Attached} regime happens when the cavity just catches up with the bottom of the mass from the droplet when it reaches the maximum radius. Thus, at higher impact energy in the \textit{Partially Attached} regime, the cavity only expands vertically as the mass from the droplet descends, forming a cylindrical cavity, where the mass from the droplet only covers part of the cavity. If the cavity obtains even higher energy, it has enough energy left when the catch up happens, such that it continues to expand radially while fully dispersing the mass from the droplet to wrap around the entire cavity, resulting in the \textit{Fully Attached} case.

\FloatBarrier

The energy balance analysis is carried out during the impact process of the suspension droplets to quantify the critical $We$ of the transitions:

\begin{equation}
    {{E}_{d,k}}+{{E}_{d,s}}={{E}_{c,p}}+{{E}_{c,s}}+{{E}_{c,k}}+{{E}_{\eta }} \label{Ebal}
\end{equation}

The left hand side captures the energy of the droplet prior to impact: kinetic energy ${{E}_{d,k}}=\pi {{\rho }_{d}}{{D}^{3}}u_{0}^{2}/12$, and surface energy ${{E}_{d,s}}=\sigma \pi {{D}^{2}}$.
During impact, the energy is converted to the pool and the cavity: cavity kinetic energy $E_{c,k}=\pi {{\rho }_{p}}{{R}_{c}^{3}}{\dot{R}_{c}^{2}}$, cavity surface energy $E_{c,s} = \sigma \pi R_{c}^{2}$, cavity potential energy $E_{c,p} = \pi {{\rho }_{p}}gR_{c}^{4}/4$, and viscous dissipation of the cornstarch suspension $E_\eta$ due to droplet deformation. Here, ${{R}_{c}}$ is the radius of the cavity and $\dot{R}_{c}$ is the radial expansion speed of the cavity.
As shown in Fig. \ref{fig_deformation}, the droplet with a low volume fraction (CST region) undergoes continuous deformation while penetrating the liquid surface, resulting in viscous dissipation. However, at higher volume fractions (BST and DST regions), the droplet initially deforms before becoming jammed. During the pre-jamming stage, energy is dissipated through plastic deformation. Once a jammed state forms, particle motion is strongly constrained by geometric packing and a force-bearing contact network, and droplet deformation is largely hindered. In this stage, the cornstarch lump exhibits negligible deformation, so additional internal viscous dissipation is neglected.
The evaluation of post-impact energy will be discussed at different characteristic times in each regime in the following sections.

\subsubsection{\label{section4A1} Origins and transitions of \textit{Liquid Lump} and \textit{Partially Attached}}

For the \textit{Liquid Lump}-\textit{Partially Attached} transition, if the cavity does not catch up with the bottom of the droplet at maximum depth, it will never catch up, such that the mass from the droplet detaches afterwards to form the \textit{Liquid Lump}, while the cavity retracts due to capillarity. Thus, the critical condition to distinguish the \textit{Liquid Lump} from the \textit{Partially Attached} regime is that the cavity just reaches the maximum depth $R_{max}$ when it catches up with the bottom of the droplet at the critical moment $t_{CST}^*$.
Since the cavity is essentially formed by the top surface of the droplet and travels at speed $u_0$, as shown in Fig. S2 of Supplemental Material~\cite{supplemental}, $t_{CST}^* = (D+R_{max})/u_0$. During this time, the bottom of the droplet travels a distance of $u_{bot}t_{CST}^* $, where $u_{bot} = 0.61u_0$ from Fig. \ref{fig_ubot}(a),  which should be equal to $R_{max}$ based on the transition criterion. Thus, the normalized critical maximum cavity depth is $\tilde R_{max}^{cr} = 1.564$.
At this critical moment, $E_{c,k}$ is negligible since the cavity reaches the maximum depth. As discussed before and shown in Fig.S4(a) of Supplemental Material~\cite{supplemental}, at such low \textit{We}, the cavity has a cone or cylindrical shape which does not strongly interact with the droplet. Thus, viscous dissipation $E_\eta$ in the mass from the droplet can be neglected.
The normalized energy balance equation Eq.(\ref{Ebal})  becomes:
\begin{equation}
    \frac{\mathrm{We}}{12}= \frac{\rho^*\tilde R_{max}^4}{2\tilde l_{c,d}^2} + (\tilde R_{\max}^2 - 1) \label{eq2}
\end{equation}

\noindent where, $\rho^* = \rho_p/{\rho_d}$ ($\rho_p$: density of the pool), $\tilde R_{max} = R_{max}/{D}$ and $\tilde l_{c,d} = \sqrt{2\sigma/\rho_dg}/D = 1.00$ is the normalized capillary length.
Plugging $\tilde R_{max}^{cr} = 1.564$  into Eq.(\ref{eq2}) gives the critical $We$ for the \textit{Liquid Lump} to \textit{Partially Attached} transition: $We_{cr,l-p} = 39.39$. The prediction of $We_{cr,l-p}$ does not involve any fitting parameters and cleanly separates the two regimes, as shown in Fig. \ref{fig_boundary_pred}(a). Note that $We_{cr,l-p}$ is independent of $\phi$, highlighting that viscous dissipation does not play an important role at such low impact speed, where the cavity does not gain enough energy to interact with the droplet to introduce strong flows.

In summary, if $We > We_{cr,l-p} = 39.39$, the cavity reaches the bottom of the droplet and its subsequent expansion partially breaks up the cornstarch lump, producing the \textit{Partially Attached} regime. Otherwise, the cavity does not have enough energy to interact with the droplet, thus the droplet remains intact and falls as a cohesive mass in the \textit{Liquid Lump} regime.

\subsubsection{\label{section4A2} Origins and transitions of \textit{Partially Attached} and \textit{Fully Attached}}

As $We$ increases, the cavity gains more energy to disperse the mass from the droplet to a thin film that wraps around the cavity. On the other hand, higher impact inertia leads to stronger internal flow in the suspension droplet, which results in higher viscous dissipation, reducing the energy transferred to the cavity, thus the mass from the droplet is not fully dispersed.
Viscous dissipation becomes important at higher $We$ for two reasons: (i) \emph{shear thickening}, which increases the effective viscosity with higher deformation rate, and (ii) \emph{geometric thinning}, as higher \(We\) produces a thinner layer and therefore a larger characteristic deformation rate (inversely proportional to the layer thickness).
The transition between the \textit{Partially Attached} and the \textit{Fully Attached} cases is thus controlled by the competition between the cavity energy and the viscous dissipation in the suspension.
If viscous dissipation in the suspension removes a substantial fraction of the initial cavity kinetic energy, cavity expansion is suppressed and the cornstarch lump cannot fully break up and attach to the cavity, leading to the \textit{Partially Attached} case. On the other hand, the cavity has so much energy compared to viscous dissipation that the cavity evolves largely unimpeded, and the impact dynamics resembles that of the Newtonian case, where the mass from the droplet is fully attached to the cavity interface (Fig.S4 of Supplemental Material~\cite{supplemental}).

The transition boundary is evaluated by comparing the initial cavity energy and the viscous dissipation as the droplet deforms. The initial cavity energy is evaluated by the energy balance model (Eq. (\ref{Ebal})) when the cavity is nascent, thus \(E_{c,s}\) and \(E_{c,p}\) are negligible. $E_{d,s}$ is negligible as \(We > 100\). Thus, the initial cavity energy scales as the kinetic energy of the droplet: $E_{c,k,0} \propto \rho_d D^3 u_0^2$. Viscous dissipation is estimated by ${E}_{\eta }={{t}^{*}}\int{{{d}^{3}}r{{\tau }_{\eta }}}\dot{\gamma }$, in which viscous stress ${{\tau }_{\eta }}=k_{\mathrm{CST}}(\phi ){{\dot{\gamma }}^{\alpha }}$
. The characteristic time scale is selected as the inertia time scale,
${{t}^{*}} = D/u_0 $, to represent the duration of droplet deformation. The shear rate $\dot{\gamma }={{u}_{0}}/D$. Thus $E_{\eta} \propto V_d k_{\mathrm{CST}}(\phi ){{u_0^{\alpha }/D}^{\alpha }}$.
Note that \(E_{c,k,0}\propto u_0^2\), which grows faster with impact velocity than \(E_\eta\) since \(\alpha<2\), which indicates that at high enough $u_0$, inertia dominates such that the droplet impact behavior resembles the Newtonian droplet impact. The transition between \textit{Partially Attached} and \textit{Fully Attached} can be obtained from the balance \(E_{c,k,0}\sim E_\eta\):

\begin{equation}
u_{0,cr,p-f} \sim \left(\frac{k_{\mathrm{CST}}(\phi)}{\rho_d\,D^{\alpha}}\right)^{\frac{1}{2-\alpha}}
\end{equation}

and

\begin{equation}
We_{cr,p-f} =\frac{\rho_d D u_{0,cr,p-f}^2}{\sigma} \propto k_{\mathrm{CST}}(\phi)^{\frac{2}{2-\alpha}}.
\end{equation}

By adopting \(\alpha = 1.5\) in the CST region ($1<\alpha<2$), the critical Weber number simplifies to \(We_{cr,p\text{-}f} \propto k_{\mathrm{CST}}(\phi)^4\). \(k_{\mathrm{CST}}(\phi)\) is extracted from rheological measurements (Fig.\ref{fig_rheolgoy}(a)) via the constitutive relation \(\tau_\eta = k_{\mathrm{CST}}(\phi) \dot{\gamma}^{1.5}\). As evidenced in Fig.S6(a) of Supplemental Material~\cite{supplemental}, \(k_{\mathrm{CST}}(\phi)\) exhibits a linear dependence on $\phi$, captured by $ k_{\mathrm{CST}}(\phi) = a\phi - b $ with best-fit parameters \(a = 0.107 \pm 0.003\) $\mathrm{Pa}\cdot \mathrm{s}^{1.5}$  and \(b = 0.037 \pm 0.001\) $\mathrm{Pa}\cdot \mathrm{s}^{1.5}$ .
To provide a semi-empirical relation for practical use,
the transition boundary is written as \(We_{cr,p-f}=c_1+c_2 k_{\mathrm{CST}}(\phi)^4\).
\(c_1\) is dimensionless and accounts for energy-loss terms not explicitly included in the simplified scaling analysis, such as the conversion of the initial droplet energy into cavity motion, wave generation, and other dissipative processes, and \(c_2\) accounts for prefactors omitted in the scaling estimate.

\begin{figure}[htb]
	\centering
	\includegraphics[width=1\linewidth]{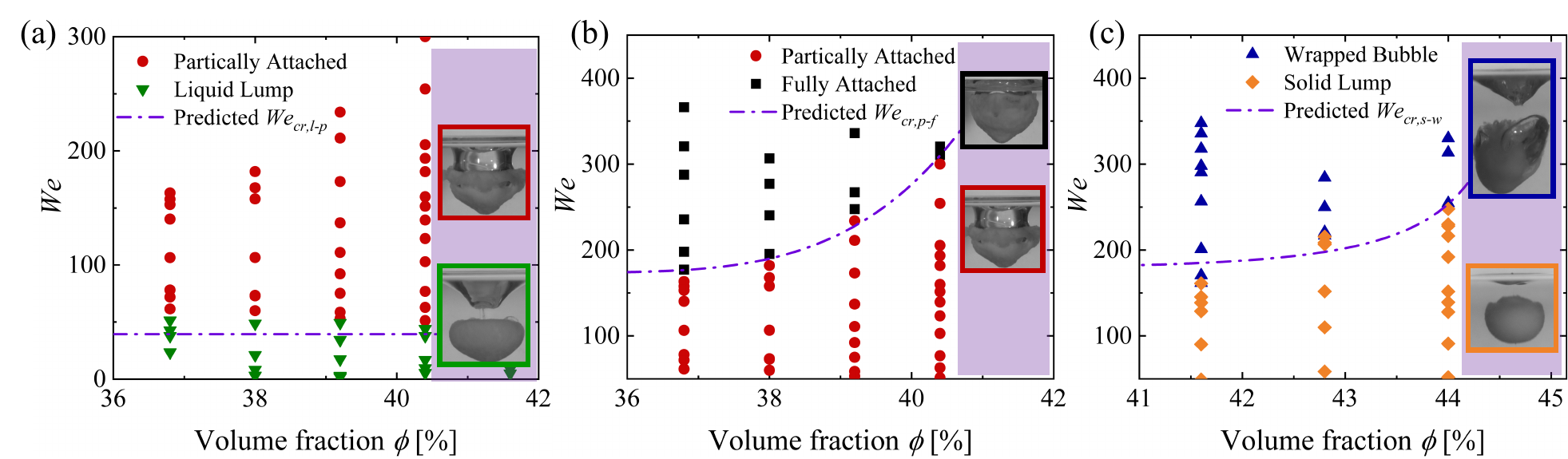}
    \caption{(a) Comparison of the predicted transition boundary $We_{cr,l-p}$ with the experimental data for \textit{Liquid Lump} and \textit{Partially Attached} cases.
    (b) Comparison of the predicted transition boundary $We_{cr,p-f}$ with the experimental data for \textit{Fully Attached} and \textit{Partially Attached} cases.
    (c) Comparison of the predicted transition boundary $We_{cr,s-w}$ with the experimental data for \textit{Wrapped Bubble} and \textit{Solid Lump} cases.} \label{fig_boundary_pred}
\end{figure}

The transition boundary fitted with $c_1 = 173.7 \pm 5.9, c_2 = 9.1 \times 10^{10} \pm 7.2 \times 10^{9}$ $(\mathrm{Pa}\cdot \mathrm{s}^{1.5})^{-4}$  is shown in Fig.\ref{fig_boundary_pred}(b).
The predicted scaling $We_{cr,p\text{-}f} \propto k_{\mathrm{CST}}(\phi)^4$ captures the experimental transition boundary well. As \(\phi\) increases, \(We_{cr,p-f}\) increases, which requires higher droplet inertia to overcome the stronger viscous dissipation for the cavity to disperse the mass from the droplet to form the \textit{Fully Attached} case.

\FloatBarrier

\subsection{\label{section4B} Bagnoldian Shear Thickening (BST) Region}

In the BST ($41.0$\%$ < \phi < 45.0$\%$ $) region with $\tau_\eta \propto \dot{\gamma}_\eta^2$, the impact phenomena: \textit{Wrapped Bubble} and \textit{Solid Lump}, differ qualitatively from those in the CST region, in terms of both the droplet and cavity dynamics. The mass from the droplet stays mostly together without being dispersed as in the CST region. The cavity dynamics does not resemble that of Newtonian droplet impact as in the CST region; instead, it either has limited expansion and is pinched off into bubbles in the \textit{Wrapped Bubble} regime, or does not occur in the \textit{Solid Lump} regime. This difference stems from the Bagnoldian shear thickening behavior $\tau_\eta \propto \dot{\gamma}_\eta^2$. As observed in similar systems with BST behavior, such as surface flow of granular materials\cite{douady2002grain} \cite{lee2025mesoscale}, step-wise deformation as quantified by the droplet height at $\phi = 42.8\%$ is observed, as shown in Fig.~\ref{fig_deformation}, further confirming the significant role of suspension rheology in impact phenomenon. Furthermore, in contrast to the CST region, the slope $\beta = u_{bot} / u_0 = 0.75$ deviates from the theoretical value of $\beta_{N,theory} = 0.59$ for Newtonian droplets impacting on liquid pools \cite{fudge2021dipping}, as shown in Fig.\ref{fig_ubot}(b).
The combination of high viscous dissipation and strong resistance between particles in the BST region stabilizes the shape of the droplet and suppresses the formation and expansion of the cavity.

To further demonstrate that the phenomena come from the BST behavior instead of a simple high viscosity effect,
droplet impact experiments with glycerol-water mixture whose viscosity matches that in the BST region are shown in Fig.S5 of Supplemental Material~\cite{supplemental}. Neither bubble entrapment nor suppressed cavity is observed in the same range of $We$, and the glycerol-water droplets continuously deform during impact, in contrast to the step-wise deformation of the suspension droplets.
In addition, the bottom section of the suspension droplets remains invariant in shape and only the upper section breaks, as shown in Fig.\ref{fig_observation}(d). This partial deformation resembles the coexisting liquid and solid phases observed for the impact of suspension droplets on solid substrates \cite{shah2022coexistence}.
Thus, the coexisting phase in the BST region leads to the \textit{Wrapped Bubble}:
the top liquid phase interacts with the pool to form the cavity, while the bottom solid phase impedes the expansion of the cavity, such that the narrowing cavity neck pinches off into bubbles as the droplet descends. At smaller $We$, the cavity does not have enough energy to pinch off, forming the \textit{Solid Lump}.
Thus, the transition criterion between \textit{Solid Lump} and \textit{Wrapped Bubble} is whether the cavity can grow large enough to pinch off when the cavity starts to evolve.

To quantify the transition boundary, the energy balance model is evaluated at the characteristic time $t_{BST}^*$, when the top surface of the droplet aligns with the undisturbed pool surface.
Since the cavity size $\tilde R_c = R_c/D$ is negligible at $t_{BST}^*$ as shown in Fig.~\ref{TransitScheme}(b), the surface energy $E_{c,s}$ and the potential energy $E_{c,p}$ of the cavity can be neglected \cite{lherm2022rayleigh}. Neglecting the droplet surface energy $E_{d,s}$ at $We>100$, the energy balance gives $E_{c,k} = E_{d,k}-E_\eta$.
Viscous dissipation ${E}_{\eta }={{t}_{BST}^{*}}\int{{{d}^{3}}r{{\tau }_{\eta }}}\dot{\gamma }$ is modeled by considering only the liquid portion of the coexisting phase for the mass transferred from the droplet.
As shown in Fig.\ref{TransitScheme}(b), droplet deformation occurs exclusively in the early stage immediately following impact, after which it ceases. Consequently, viscous dissipation is attributed solely to the deformation stage and ceases once the deformation is arrested by particle jamming.
This motivates the introduction of the dissipation fraction inside the droplet $\theta$, which gives $E_{c,k} = E_{d,k}-\theta {{t}_{BST}^{*}}\int{{{d}^{3}}r{{\tau }_{\eta }}}\dot{\gamma }$. $\theta$ serves as a fitting parameter in this equation.
In addition, $k_{BST}(\phi) = 5.3\times {{10}^{-9}}(1-\phi/0.52)^{-7.69}$ $\mathrm{Pa}\cdot \mathrm{s}^{2}$  is fitted from the rheology measurement as shown in Fig.S6(b) of Supplemental Material~\cite{supplemental}.

The transition criterion is given by whether there is enough remaining kinetic energy $E_{c,k}=E_{c,k,cr}$ for the cavity to pinch off to form the \textit{Wrapped Bubble}. Note that unlike the transition from \textit{Partially Attached} to \textit{Fully Attached} in the CST region, where the expansion of the cavity competes against viscous dissipation throughout the cavity development, the viscous dissipation in the BST region happens at the initial stage to reduce the energy transferred to the cavity, and the cavity expansion is confined by the jammed solid phase. Due to the challenge in theoretically predicting $E_{c,k,cr}$, it is absorbed into the fitting parameters when carrying out the energy balance to find the semi-empirical transition  $We_{cr,s-w}$:

\begin{equation}
We_{cr,s-w}=\frac{c_3}{\left(1/12 - c_4\,k_{\mathrm{BST}}(\phi)\right)}\label{eq109}
\end{equation}

\noindent shown as the dash-dotted line in Fig.~\ref{fig_boundary_pred}(c) with \(c_3=177.6 \pm 11.4\) and \(c_4=31.7 \pm 6.8 \) $(\mathrm{Pa}\cdot \mathrm{s}^{2})^{-1}$.
Here, $c_3$ accounts for the finite cavity energy threshold required for bubble pinch-off, whereas $c_4$ absorbs the dissipation fraction $\theta$, geometric factors, and other prefactors omitted in the scaling estimate.
The fitted boundary agrees well with the experimental data and captures the increase of \(We_{cr,s-w}\) with \(\phi\): higher \(\phi\) leads to stronger viscous dissipation, and therefore requires larger inertia to generate a cavity large enough to pinch off to form the \textit{Wrapped Bubble}.

\subsection{\label{section4C} Discontinuous Shear Thickening (DST) Region}

At high volume fraction ($\phi > 45.0\% $), the \textit{Wrapped Bubble} regime vanishes and only \textit{Solid Lump} occurs. In this DST dominated region, particle motion is restricted, causing the droplet to behave like a solid. As shown in Fig.~\ref{fig_deformation} for $\phi = 47.5\%$, after an initial deformation, the height of the droplet reaches a plateau, indicating impact-induced jamming \cite{liu1998jamming, bi2011jamming, waitukaitis2012impact}.
Further evidence of impact-induced jamming comes from cavity suppression, as shown in Fig.\ref{fig_observation}(e), which resembles the case of a hydrophilic solid sphere impacting a liquid pool \cite{duez2007making}. Therefore, although the droplet remains liquid prior to impact, it solidifies during impact.
Further discussion of the phase transition during the impact will be detailed in our future work.

From the energy point of view, the cavity energy $E_c$ scales as the droplet kinetic energy $E_{d,k}=\pi {{\rho }_{d}}{{D}^{3}}u_{0}^{2}/12$, where the viscous dissipation $E_\eta \propto V_dk_{DST}(\phi ){{u_0^{\alpha }/D}^{\alpha }}$ with $\alpha > 2$. Thus, with increasing $u_0$, $E_{\eta}$ increases much faster than $E_c$. Therefore, in the DST region, even at high $u_0$, the large $E_c$ gained from droplet inertia can still be dissipated by viscous dissipation $E_\eta$ due to strong shear thickening, leading to diminished cavity, such that the \textit{Wrapped Bubble} disappears in the DST region.

\section{\label{section6} Conclusion}

\begin{figure}[htb]
	\centering
    \includegraphics[height = 0.5\textwidth]{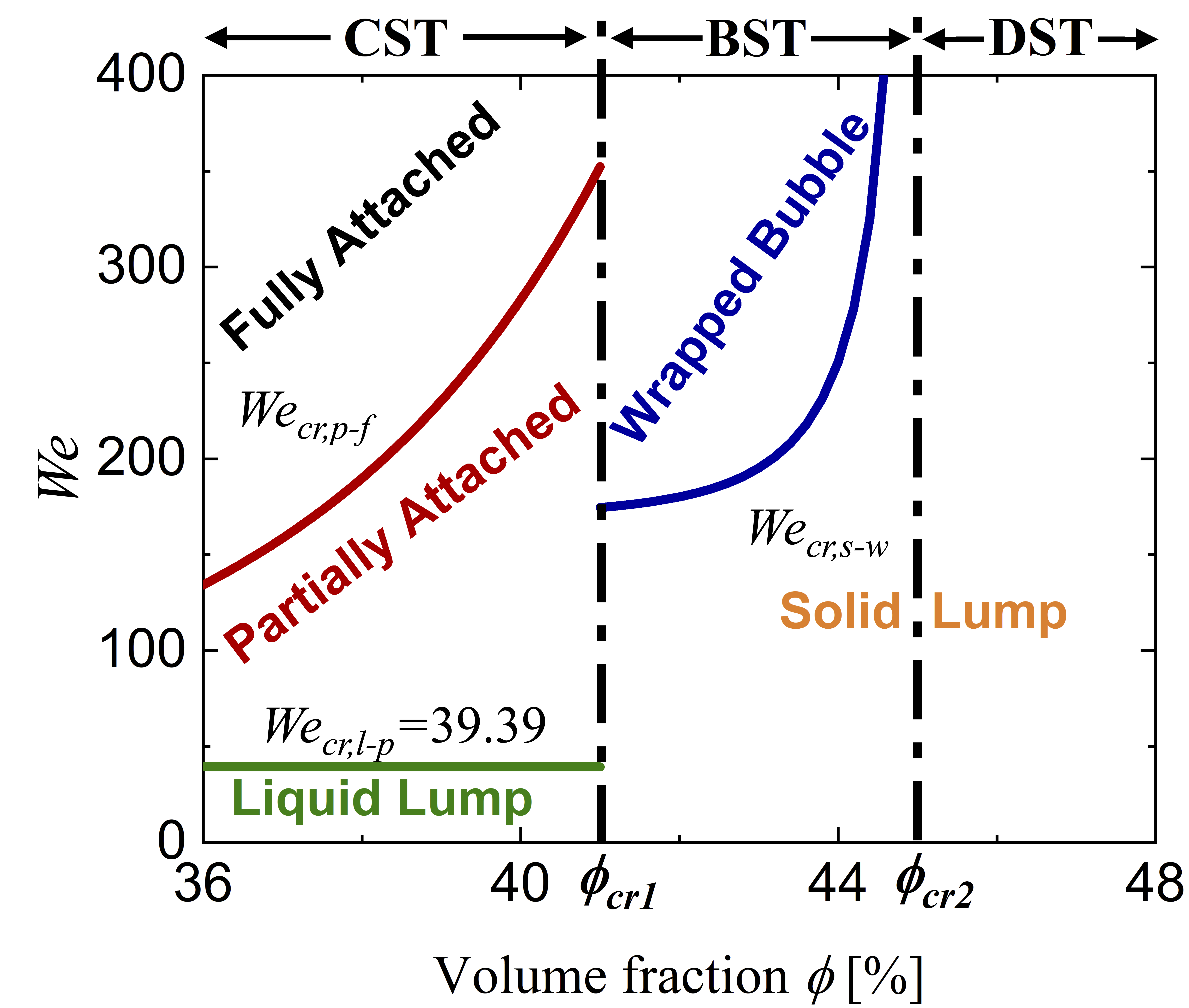}
    \caption{Regime map with predicted Weber number for the transition boundaries of cornstarch droplet impacting on a liquid pool.
    $We_{cr,l-p} = 39.39 $;
    $We_{cr,p-f} = 173.7 + 9.1 \times 10^{10} k_{CST}(\phi )^{4} $, where $k_{CST}(\phi )=0.107\phi -0.037$ $\mathrm{Pa}\cdot \mathrm{s}^{1.5}$;
    $We_{cr,s-w} = 177.6/(1/12 -  31.7 k_{BST}(\phi ))$, where $k_{BST}(\phi )=5.3 \times 10^{-9}(1-\phi/0.52)^{-7.69}$ $\mathrm{Pa}\cdot \mathrm{s}^{2}$. The comparison between the experimental data and the predicted transition boundaries is shown in Fig. \ref{fig_boundary_pred}.
     } \label{fig_summary}
\end{figure}

In this work, we experimentally investigated the impact dynamics of cornstarch suspension droplets on a deep water pool.
We identified a regime map in the $We-\phi$ space consisting of five distinct impact regimes
(Fig.\ref{fig_regime}) and proposed theoretical predictions for the transition boundaries as summarized in Fig.~\ref{fig_summary}. Rheology tests of the cornstarch suspension revealed that the two critical volume fractions $\phi_{cr,1}\approx41\%$ and $\phi_{cr,2}\approx45\%$, which divide the regime map into three regions, are controlled by the transition of the suspension behavior from CST to BST to DST, based on the value of $\alpha$ in the power law $\tau_\eta \propto \dot{\gamma}^\alpha$. In the CST region ($1< \alpha < 2$, $\phi < \phi_{cr,1}$)
\textit{Liquid Lump}, \textit{Partially Attached} and \textit{Fully Attached} occur, where the cavity dynamics closely resembles that of Newtonian droplet impact, and the weak viscosity variation modifies how particles are dispersed by the cavity expansion.
The BST region ($\alpha = 2$, $\phi_{cr,1}< \phi < \phi_{cr,2}$)
contains \textit{Wrapped Bubble} and \textit{Solid Lump} regimes, both fundamentally different from the impact dynamics of Newtonian droplets. The coexistence of the liquid and solid phases in the droplet during impact leads to the \textit{Wrapped Bubble}, where at low $We$, \textit{Solid Lump} occurs due to the jamming of the particles.
In the DST region ($\alpha > 2$, $\phi > \phi_{cr,2}$), only \textit{Solid Lump} occurs, where impact-induced jamming occurs at all $We$, causing the droplet to behave like a solid sphere.

\FloatBarrier

We further quantified the critical Weber number for each transition boundary in the CST and BST regions (summarized in Fig.~\ref{fig_summary}) by carrying out energy balance analysis during the droplet impact, where the coupling between the cavity and non-Newtonian behavior of the suspension plays a critical role.
The scaling predictions of the transition boundaries agree well with the experimental data by capturing the underlying physics. We further provided semi-empirical expressions that quantitatively describe the transition boundaries for practical usage and offer a systematic foundation for further studies of suspension droplets impact on liquid surfaces.
Our study fills a research gap on the impact dynamics of suspension droplets on a liquid pool and deepens our understanding of the interplay between impact dynamics, cavity evolution, and rheological properties, providing practical guidance for engineering applications such as controlling cavity formation, improving precision in 3D printing, and optimizing medical storage.

\section*{Data Availability}

Most data supporting the findings of this article are presented in the figures. Data are available from the corresponding author upon reasonable request.

\begin{acknowledgments}
The authors acknowledge NSF funding support under awards \#2227985. The authors thank Professor Sara
Hashmi and Dr. Saeed Alborzi for their support with rheological measurements.
\end{acknowledgments}

\bibliography{apssamp_f}

@PREAMBLE{
 "\providecommand{\noopsort}[1]{}" 
 # "\providecommand{\singleletter}[1]{#1}%" 
}

@misc{supplemental,
  title = {See Supplemental Material at [journals.aps.org/prfluids/accepted/10.1103/fxw2-38g4] for details on the volume-fraction calculation, image measurements, regime classification, reference impact dynamics of water droplet and glycerol-water droplet, and rheological fitting procedure.},
  note = {}
}

@article{han2019dynamic,
  title={Dynamic jamming of dense suspensions under tilted impact},
  author={Han, Endao and Zhao, Liang and Van Ha, Nigel and Hsieh, S Tonia and Szyld, Daniel B and Jaeger, Heinrich M},
  journal={Physical Review Fluids},
  volume={4},
  number={6},
  pages={063304},
  year={2019},
  publisher={APS}
}

@article{bertola2015impact,
  title={Impact of concentrated colloidal suspension drops on solid surfaces},
  author={Bertola, Volfango and Haw, Mark D},
  journal={Powder technology},
  volume={270},
  pages={412--417},
  year={2015},
  publisher={Elsevier}
}

@article{cao2024regulating,
  title={Regulating droplet impact dynamics of nanoparticle suspension: Phenomena, mechanisms, and implications},
  author={Cao, Hao and Hu, Ran and Chen, Yi-Feng and Gui, Chengqun and Yang, Zhibing and others},
  journal={Physics of Fluids},
  volume={36},
  number={1},
  year={2024},
  publisher={AIP Publishing}
}

@article{douady2002grain,
  title={From a grain to avalanches: on the physics of granular surface flows},
  author={Douady, St{\'e}phane and Andreotti, Bruno and Daerr, Adrian and Clad{\'e}, Pierre},
  journal={Comptes Rendus Physique},
  volume={3},
  number={2},
  pages={177--186},
  year={2002},
  publisher={Elsevier}
}

@article{bi2011jamming,
  title={Jamming by shear},
  author={Bi, Dapeng and Zhang, Jie and Chakraborty, Bulbul and Behringer, Robert P},
  journal={Nature},
  volume={480},
  number={7377},
  pages={355--358},
  year={2011},
  publisher={Nature Publishing Group UK London}
}

@article{morris2020toward,
  title={Toward a fluid mechanics of suspensions},
  author={Morris, Jeffrey F},
  journal={Phys. Rev. Fluids.},
  volume={5},
  number={11},
  pages={110519},
  year={2020},
  publisher={APS}
}

@article{duez2007making,
  title={Making a splash with water repellency},
  author={Duez, Cyril and Ybert, Christophe and Clanet, Christophe and Bocquet, Lyderic},
  journal={Nat. Phys.},
  volume={3},
  number={3},
  pages={180--183},
  year={2007},
  publisher={Nature Publishing Group UK London}
}

@article{morris2020shear,
  title={Shear thickening of concentrated suspensions: Recent developments and relation to other phenomena},
  author={Morris, Jeffrey F},
  journal={Annu. Rev. Fluid Mech.},
  volume={52},
  pages={121--144},
  year={2020},
  publisher={Annual Reviews}
}

@article{yarin2006drop,
  title={Drop impact dynamics: splashing, spreading, receding, bouncing.},
  author={Yarin, Alexander L},
  journal={Annu. Rev. Fluid Mech.},
  volume={38},
  pages={159--192},
  year={2006},
  publisher={Annual Reviews}
}

@article{rein1993phenomena,
  title={Phenomena of liquid drop impact on solid and liquid surfaces},
  author={Rein, Martin},
  journal={Fluid. Dyn. Res.},
  volume={12},
  number={2},
  pages={61--93},
  year={1993},
  publisher={Elsevier}
}

@article{fudge2021dipping,
  title={Dipping into a new pool: The interface dynamics of drops impacting onto a different liquid},
  author={Fudge, Ben D and Cimpeanu, Radu and Castrej{\'o}n-Pita, Alfonso A},
  journal={Physical Review E},
  volume={104},
  number={6},
  pages={065102},
  year={2021},
  publisher={APS}
}

@article{lherm2022rayleigh,
  title={Rayleigh--Taylor instability in impact cratering experiments},
  author={Lherm, Victor and Deguen, Renaud and Alboussi{\`e}re, Thierry and Landeau, Maylis},
  journal={Journal of Fluid Mechanics},
  volume={937},
  pages={A20},
  year={2022},
  publisher={Cambridge University Press}
}

@article{tang2019bouncing,
  title={Bouncing drop on liquid film: Dynamics of interfacial gas layer},
  author={Tang, Xiaoyu and Saha, Abhishek and Law, Chung K and Sun, Chao},
  journal={Phys. Fluids},
  volume={31},
  number={1},
  pages={013304},
  year={2019},
  publisher={AIP Publishing LLC}
}

@article{tang2018bouncing,
  title={Bouncing-to-merging transition in drop impact on liquid film: Role of liquid viscosity},
  author={Tang, Xiaoyu and Saha, Abhishek and Law, Chung K and Sun, Chao},
  journal={Langmuir},
  volume={34},
  number={8},
  pages={2654--2662},
  year={2018},
  publisher={ACS Publications}
}

@article{truscott2014water,
  title={Water entry of projectiles},
  author={Truscott, Tadd T and Epps, Brenden P and Belden, Jesse},
  journal={Annu. Rev. Fluid Mech.},
  volume={46},
  pages={355--378},
  year={2014},
  publisher={Annual Reviews}
}

@article{aristoff2010water,
  title={The water entry of decelerating spheres},
  author={Aristoff, Jeffrey M and Truscott, Tadd T and Techet, Alexandra H and Bush, John WM},
  journal={Phys. Fluids},
  volume={22},
  number={3},
  pages={032102},
  year={2010},
  publisher={American Institute of Physics}
}

@article{tang2019spreading,
  title={Spreading and oscillation dynamics of drop impacting liquid film},
  author={Tang, Xiaoyu and Saha, Abhishek and Sun, Chao and Law, Chung K},
  journal={J. Fluid Mech},
  volume={881},
  pages={859--871},
  year={2019},
  publisher={Cambridge University Press}
}

@book{mewis2012colloidal,
  title={Colloidal suspension rheology},
  author={Mewis, Jan and Wagner, Norman J},
  year={2012},
  publisher={Cambridge university press}
}

@article{liu1998jamming,
  title={Jamming is not just cool any more},
  author={Liu, Andrea J and Nagel, Sidney R},
  journal={Nature},
  volume={396},
  number={6706},
  pages={21--22},
  year={1998},
  publisher={Nature Publishing Group UK London}
}

@article{waitukaitis2012impact,
  title={Impact-activated solidification of dense suspensions via dynamic jamming fronts},
  author={Waitukaitis, Scott R and Jaeger, Heinrich M},
  journal={Nature},
  volume={487},
  number={7406},
  pages={205--209},
  year={2012},
  publisher={Nature Publishing Group UK London}
}

@article{han2016high,
  title={High-speed ultrasound imaging in dense suspensions reveals impact-activated solidification due to dynamic shear jamming},
  author={Han, Endao and Peters, Ivo R and Jaeger, Heinrich M},
  journal={Nat. Commun.},
  volume={7},
  number={1},
  pages={12243},
  year={2016},
  publisher={Nature Publishing Group UK London}
}

@article{Boyer2016Drop,
  title={Drop impact of shear thickening liquids},
  author={ Boyer, F.  and  Snoeijer, J. H.  and  Dijksman, J. F.  and  Lohse, D. },
  journal={Phys. Rev. Fluid},
  volume={1},
  number={1},
  year={2016},
}

@article{shah2022coexistence,
  title={Coexistence of solid and liquid phases in shear jammed colloidal drops},
  author={Shah, Phalguni and Arora, Srishti and Driscoll, Michelle M},
  journal={Commun. Phys.},
  volume={5},
  number={1},
  pages={222},
  year={2022},
  publisher={Nature Publishing Group UK London}
}

@article{Han2017Measuring,
  title={Measuring the porosity and compressibility of liquid-suspended porous particles using ultrasound},
  author={ Han, Endao  and  Ha, Nigel Van  and  Jaeger, Heinrich M },
  journal={Soft Matter},
  volume={13},
  number={19},
  year={2017},
}

@article{arashiro1999use,
  title={Use of the pendant drop method to measure interfacial tension between molten polymers},
  author={Arashiro, Emerson Y and Demarquette, Nicole R},
  journal={Mater. Res.},
  volume={2},
  pages={23--32},
  year={1999},
  publisher={SciELO Brasil}
}

@article{Fall2012Shear,
  title={Shear thickening of cornstarch suspensions},
  author={ Fall, A.  and  Bertrand, F.  and  Ovarlez, G.  and  Bonn, D. },
  journal={J. Rheol.},
  volume={56},
  number={3},
  pages={145-150},
  year={2012},
}

@article{shah2024drop,
  title={Drop impact dynamics of complex fluids: A review},
  author={Shah, Phalguni and Driscoll, Michelle M},
  journal={Soft Matter},
  volume={20},
  number={25},
  pages={4839--4858},
  year={2024},
  publisher={Royal Society of Chemistry}
}

@article{Tanner2018Review,
  title={Review Article: Aspects of non-colloidal suspension rheology},
  author={ Tanner, R. I. },
  journal={Phys. Fluids},
  volume={30},
  number={10},
  year={2018},
}

@article{Brown2014Shear,
  title={Shear thickening in concentrated suspensions: phenomenology, mechanisms and relations to jamming},
  author={ Brown, E. and Jaeger, H. M. },
  journal={Rep. Prog. Phys.},
  volume={77},
  number={4},
  pages={046602},
  year={2014},
}

@article{Stauffer1965The,
  title={The Measurement of Surface Tension by the Pendant Drop Technique},
  author={ Stauffer, C. E. },
  journal={J. Phys. Chem.},
  volume={69},
  year={1965},
}

@article{pan2007dynamics,
  title={Dynamics of droplet--film collision},
  author={Pan, Kuo Long and Law, Chung K},
  journal={J. Fluid Mech.},
  volume={587},
  pages={1--22},
  year={2007},
  publisher={Cambridge University Press}
}

@article{che2018impact,
  title={Impact of droplets on immiscible liquid films},
  author={Che, Zhizhao and Matar, Omar K},
  journal={Soft Matter},
  volume={14},
  number={9},
  pages={1540--1551},
  year={2018},
  publisher={Royal Society of Chemistry}
}

@article{kim2020raindrop,
  title={How a raindrop gets shattered on biological surfaces},
  author={Kim, Seungho and Wu, Zixuan and Esmaili, Ehsan and Dombroskie, Jason J and Jung, Sunghwan},
  journal={Proc. Natl. Acad. Sci. U.S.A.},
  volume={117},
  number={25},
  pages={13901--13907},
  year={2020},
  publisher={National Acad Sciences}
}

@article{gart2015droplet,
  title={Droplet impacting a cantilever: A leaf-raindrop system},
  author={Gart, Sean and Mates, Joseph E and Megaridis, Constantine M and Jung, Sunghwan},
  journal={Phys. Rev. Appl.},
  volume={3},
  number={4},
  pages={044019},
  year={2015},
  publisher={APS}
}

@article{bagchi2021penetration,
  title={Penetration and secondary atomization of droplets impacted on wet facemasks},
  author={Bagchi, Sombuddha and Basu, Saptarshi and Chaudhuri, Swetaprovo and Saha, Abhishek},
  journal={Phys. Rev. Fluid},
  volume={6},
  number={11},
  pages={110510},
  year={2021},
  publisher={APS}
}

@article{dressaire2016drop,
  title={Drop impact on a flexible fiber},
  author={Dressaire, Emilie and Sauret, Alban and Boulogne, Fran{\c{c}}ois and Stone, Howard A},
  journal={Soft Matter},
  volume={12},
  number={1},
  pages={200--208},
  year={2016},
  publisher={R. Soc. Chem. paper.}
}

@article{pack2017failure,
  title={Failure mechanisms of air entrainment in drop impact on lubricated surfaces},
  author={Pack, Min and Hu, Han and Kim, D and Zheng, Zhong and Stone, HA and Sun, Ying},
  journal={Soft Matter},
  volume={13},
  number={12},
  pages={2402--2409},
  year={2017},
  publisher={Royal Society of Chemistry}
}

@article{tsai2009drop,
  title={Drop impact upon micro-and nanostructured superhydrophobic surfaces},
  author={Tsai, Peichun and Pacheco, Sergio and Pirat, Christophe and Lefferts, Leon and Lohse, Detlef},
  journal={Langmuir},
  volume={25},
  number={20},
  pages={12293--12298},
  year={2009},
  publisher={ACS Publications}
}

@article{guemas2012drop,
  title={Drop impact experiments of non-Newtonian liquids on micro-structured surfaces},
  author={Gu{\'e}mas, Marine and Mar{\'\i}n, {\'A}lvaro G and Lohse, Detlef},
  journal={Soft Matter},
  volume={8},
  number={41},
  pages={10725--10731},
  year={2012},
  publisher={Royal Society of Chemistry}
}

@article{jain2019deep,
  title={Deep pool water-impacts of viscous oil droplets},
  author={Jain, Utkarsh and Jalaal, Maziyar and Lohse, Detlef and van der Meer, Devaraj},
  journal={Soft Matter},
  volume={15},
  number={23},
  pages={4629--4638},
  year={2019},
  publisher={Royal Society of Chemistry}
}

@article{richard2002contact,
  title={Contact time of a bouncing drop},
  author={Richard, Denis and Clanet, Christophe and Qu{\'e}r{\'e}, David},
  journal={Nature},
  volume={417},
  number={6891},
  pages={811--811},
  year={2002},
  publisher={Nature Publishing Group UK London}
}

@article{de2015wettability,
  title={Wettability-independent bouncing on flat surfaces mediated by thin air films},
  author={De Ruiter, Jolet and Lagraauw, Rudy and Van Den Ende, Dirk and Mugele, Frieder},
  journal={Nat. Phys.},
  volume={11},
  number={1},
  pages={48--53},
  year={2015},
  publisher={Nature Publishing Group UK London}
}

@article{kolinski2014drops,
  title={Drops can bounce from perfectly hydrophilic surfaces},
  author={Kolinski, John M and Mahadevan, L and Rubinstein, Shmuel M},
  journal={EPL},
  volume={108},
  number={2},
  pages={24001},
  year={2014},
  publisher={IOP Publishing}
}

@article{zou2011experimental,
  title={Experimental study of a drop bouncing on a liquid surface},
  author={Zou, Jun and Wang, Peng Fei and Zhang, Ting Rong and Fu, Xin and Ruan, Xiaodong},
  journal={Phys. Fluids},
  volume={23},
  number={4},
  pages={044101},
  year={2011},
  publisher={American Institute of Physics}
}

@article{he2021drop,
  title={Drop bouncing dynamics on ultrathin films},
  author={He, Ziwen and Tran, Huy and Pack, Min Y},
  journal={Langmuir},
  volume={37},
  number={33},
  pages={10135--10142},
  year={2021},
  publisher={ACS Publications}
}

@article{lee2016universal,
  title={Universal rescaling of drop impact on smooth and rough surfaces},
  author={Lee, JB and Laan, Nick and de Bruin, Karla G and Skantzaris, G and Shahidzadeh, Noushine and Derome, Dominique and Carmeliet, J and Bonn, Daniel},
  journal={J. Fluid Mech.},
  volume={786},
  pages={R4},
  year={2016},
  publisher={Cambridge University Press}
}

@article{lagubeau2012spreading,
  title={Spreading dynamics of drop impacts},
  author={Lagubeau, Guillaume and Fontelos, Marco A and Josserand, Christophe and Maurel, Agn{\`e}s and Pagneux, Vincent and Petitjeans, Philippe},
  journal={J. Fluid Mech.},
  volume={713},
  pages={50--60},
  year={2012},
  publisher={Cambridge University Press}
}

@article{roisman2002normal,
  title={Normal impact of a liquid drop on a dry surface: model for spreading and receding},
  author={Roisman, Ilia V and Rioboo, Romain and Tropea, Cameron},
  journal={Philos. Trans. Royal Soc. A},
  volume={458},
  number={2022},
  pages={1411--1430},
  year={2002},
  publisher={The Royal Society}
}

@article{laan2014maximum,
  title={Maximum diameter of impacting liquid droplets},
  author={Laan, Nick and de Bruin, Karla G and Bartolo, Denis and Josserand, Christophe and Bonn, Daniel},
  journal={Phys. Rev. Appl.},
  volume={2},
  number={4},
  pages={044018},
  year={2014},
  publisher={APS}
}

@article{wildeman2016spreading,
  title={On the spreading of impacting drops},
  author={Wildeman, Sander and Visser, Claas Willem and Sun, Chao and Lohse, Detlef},
  journal={J. Fluid Mech.},
  volume={805},
  pages={636--655},
  year={2016},
  publisher={Cambridge University Press}
}

@article{peng2021droplet,
  title={Droplet splashing during the impact on liquid pools of shear-thinning fluids with yield stress},
  author={Peng, Xiaoyun and Wang, Tianyou and Sun, Kai and Che, Zhizhao},
  journal={Phys. Fluids},
  volume={33},
  number={11},
  pages={112106},
  year={2021},
  publisher={AIP Publishing LLC}
}

@article{juarez2012splash,
  title={Splash control of drop impacts with geometric targets},
  author={Juarez, Gabriel and Gastopoulos, Thomai and Zhang, Yibin and Siegel, Michael L and Arratia, Paulo E},
  journal={Phys. Rev. E},
  volume={85},
  number={2},
  pages={026319},
  year={2012},
  publisher={APS}
}

@article{yang2021experimental,
  title={Experimental study on droplet splash and receding breakup on a smooth surface at atmospheric pressure},
  author={Yang, Lei and Li, Zhonghong and Yang, Tao and Chi, Yicheng and Zhang, Peng},
  journal={Langmuir},
  volume={37},
  number={36},
  pages={10838--10848},
  year={2021},
  publisher={ACS Publications}
}

@article{murphy2015splash,
  title={Splash behaviour and oily marine aerosol production by raindrops impacting oil slicks},
  author={Murphy, David W and Li, Cheng and d’Albignac, Vincent and Morra, David and Katz, Joseph},
  journal={J. Fluid Mech.},
  volume={780},
  pages={536--577},
  year={2015},
  publisher={Cambridge University Press}
}

@article{zhang2010wavelength,
  title={Wavelength selection in the crown splash},
  author={Zhang, Li V and Brunet, Philippe and Eggers, Jens and Deegan, Robert D},
  journal={Phys. Fluids},
  volume={22},
  number={12},
  pages={122105},
  year={2010},
  publisher={American Institute of Physics}
}

@article{xu2005drop,
  title={Drop splashing on a dry smooth surface},
  author={Xu, Lei and Zhang, Wendy W and Nagel, Sidney R},
  journal={Phys. Rev. Lett.},
  volume={94},
  number={18},
  pages={184505},
  year={2005},
  publisher={APS}
}

@article{mani2010events,
  title={Events before droplet splashing on a solid surface},
  author={Mani, Madhav and Mandre, Shreyas and Brenner, Michael P},
  journal={J. Fluid Mech.},
  volume={647},
  pages={163--185},
  year={2010},
  publisher={Cambridge University Press}
}

@article{bagnold1954experiments,
  title={Experiments on a gravity-free dispersion of large solid spheres in a Newtonian fluid under shear},
  author={Bagnold, Ralph Alger},
  journal={Proceedings of the Royal Society of London. Series A. Mathematical and Physical Sciences},
  volume={225},
  number={1160},
  pages={49--63},
  year={1954},
  publisher={The Royal Society London}
}

@article{lee2025mesoscale,
  title={Mesoscale avalanche size underpins the rheology of granular yielding},
  author={Lee, Keng-Lin and Yeh, Ting-Yin},
  journal={Proceedings of the National Academy of Sciences},
  volume={122},
  number={40},
  pages={e2516426122},
  year={2025},
  publisher={National Academy of Sciences}
}

@article{marschall2003cavitation,
  title={Cavitation inception by almost spherical solid particles in water},
  author={Marschall, HB and M{\o}rch, Knud Aage and Keller, AP and Kjeldsen, M},
  journal={Phys. Fluids},
  volume={15},
  number={2},
  pages={545--553},
  year={2003},
  publisher={American Institute of Physics}
}

@article{liu2011extension,
  title={Extension of the St{\"o}ber method to the preparation of monodisperse resorcinol--formaldehyde resin polymer and carbon spheres},
  author={Liu, Jian and Qiao, Shi Zhang and Liu, Hao and Chen, Jun and Orpe, Ajay and Zhao, Dongyuan and Lu, Gao Qing},
  journal={Angewandte Chemie},
  volume={123},
  number={26},
  pages={6069--6073},
  year={2011},
  publisher={WILEY-VCH Verlag Weinheim}
}

\end{document}

\section*{Supplementary Material}

\section{\label{Sup1} Quantification of volume fraction of cornstarch suspension}

Cornstarch suspension is made by weighing the cornstarch particles, while most suspension rheology models are based on the volume fraction. The calculation of volume fraction is not trivial since cornstarch particles are porous such that ambient water vapor can occupy their pores. To account for this, we adopted the correction methods by Han et al.~\cite{Han2017Measuring, han2016high} that convert the measured mass (weight) fraction into the particle volume fraction $\phi$. Specifically, $\phi$ is obtained from the following two equations:

\begin{equation} \label{eq:appA1}
{{\phi }_{v}}=\frac{(1-\beta ){{m}_{c}}/{{\rho }_{c}}}{(1-\beta ){{m}_{c}}/{{\rho }_{c}}+{{m}_{s}}/{{\rho }_{s}}+\beta {{m}_{c}}/{{\rho }_{c}}}
\end{equation}

\begin{equation} \label{eq:appA2}
\phi =\frac{{\phi }_{v}}{1-\psi}
\end{equation}

\noindent where $m$ denotes mass and $\rho$ density, with subscripts $c$ and $s$ referring to cornstarch and solvent, respectively. 
Here, $\phi_v$ is the apparent volume fraction of cornstarch particles calculated from mass measurements, $\psi$ is the porosity of the cornstarch granules representing the fraction of internal pore space accessible to the solvent and $\beta$ is the moisture content of cornstarch powder under ambient conditions. 
Han et al.~\cite{Han2017Measuring, han2016high} recommended $\psi =0.3, \beta =0.11$ based on prior characterization studies. Therefore, based on the weight fraction between the particles and the solvent $m_c / m_s$, the volume fraction $\phi$ can be calculated. In our experiments, we varied the weight ratio $m_c / m_s$ between cornstarch particles and the Cesium Chloride (CsCl) solution to obtain the desired $\phi$.

\section{\label{Sup0}Measurements}

The measurement procedure is divided into two stages relative to the moment of impact, defined as $t = 0$ s, when the droplet first contacts the liquid pool surface.

\textbf{Before impact} ($t < 0$ s), the droplet falls freely towards the pool. In each frame, the projected area $A$ of the droplet is extracted from the image and used to compute the equivalent spherical diameter $D = \sqrt{4A/\pi}$. 
The variation of $D$ with volume fraction $\phi$ is shown in Fig.S1(a).
The vertical displacement of the droplet centroid $y$ is tracked over time, shown in Fig.S1(b), and the impact velocity $u_0$ is determined by fitting a linear relationship to the displacement-time data.

\textbf{After impact} ($t > 0$ s), the droplet merges with and penetrates the liquid pool.
%, and the free surfaces deform. 
The vertical displacements of the top point and bottom point of the merged droplet are tracked independently over time. 
As illustrated in Fig.~\ref{fig_ubot_utop_mearsure}, linear fits are applied to each displacement-time relation to extract characteristic velocities $u_\mathrm{top}$ and $u_\mathrm{bot}$, respectively.
The result shows that $u_{top}$ and $u_{bot}$ are constants and $u_{top} < u_{bot}$, which signifies that the droplet undergoes a vertical deformation during penetration.

\begin{figure}[htb]
	\centering
	\includegraphics[width = 0.95\linewidth]{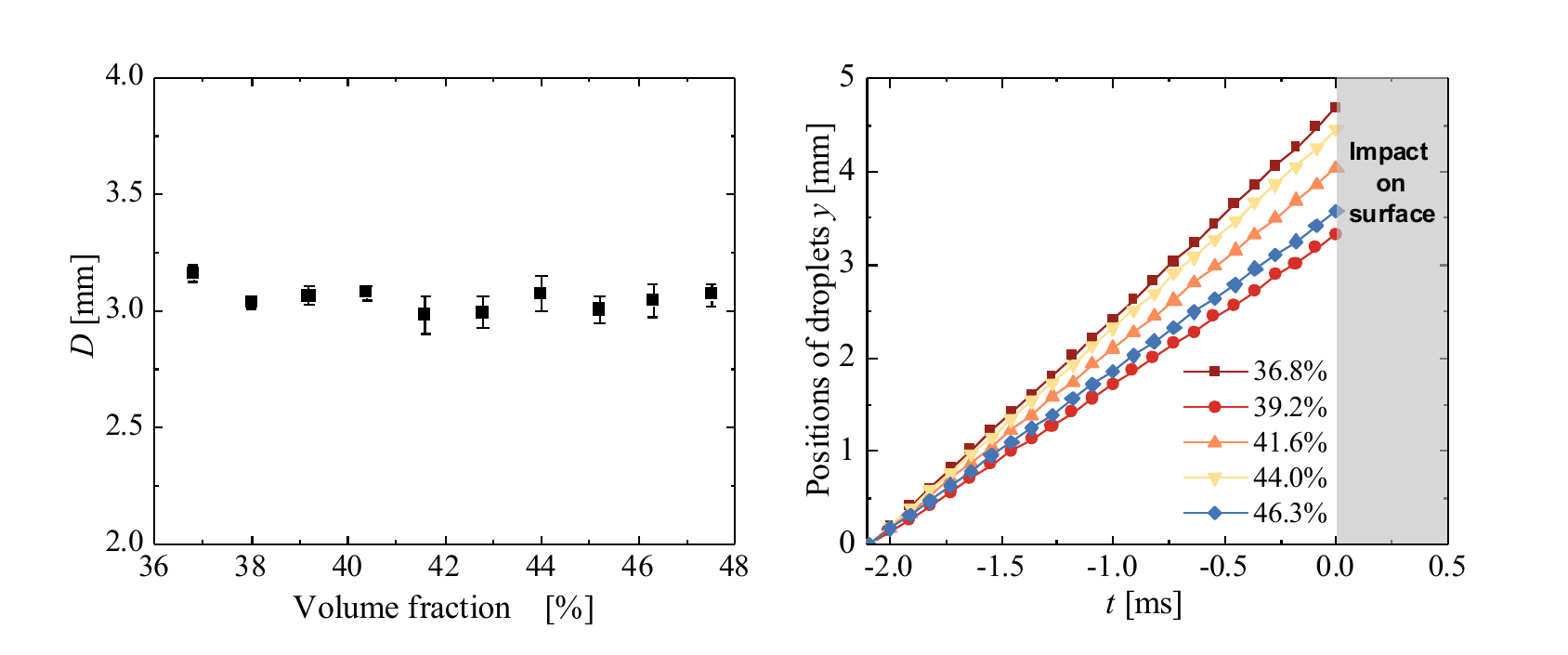}
    \caption{(a) Droplet diameter vs particle volume fraction $\phi$. (b) Pre-impact centroid displacements for different $\phi$. The position $y = 0$ corresponds to the pool surface.} \label{fig_measurement}
\end{figure}
\FloatBarrier

\begin{figure}[htb]
	\centering
    \includegraphics[width=0.6\linewidth]{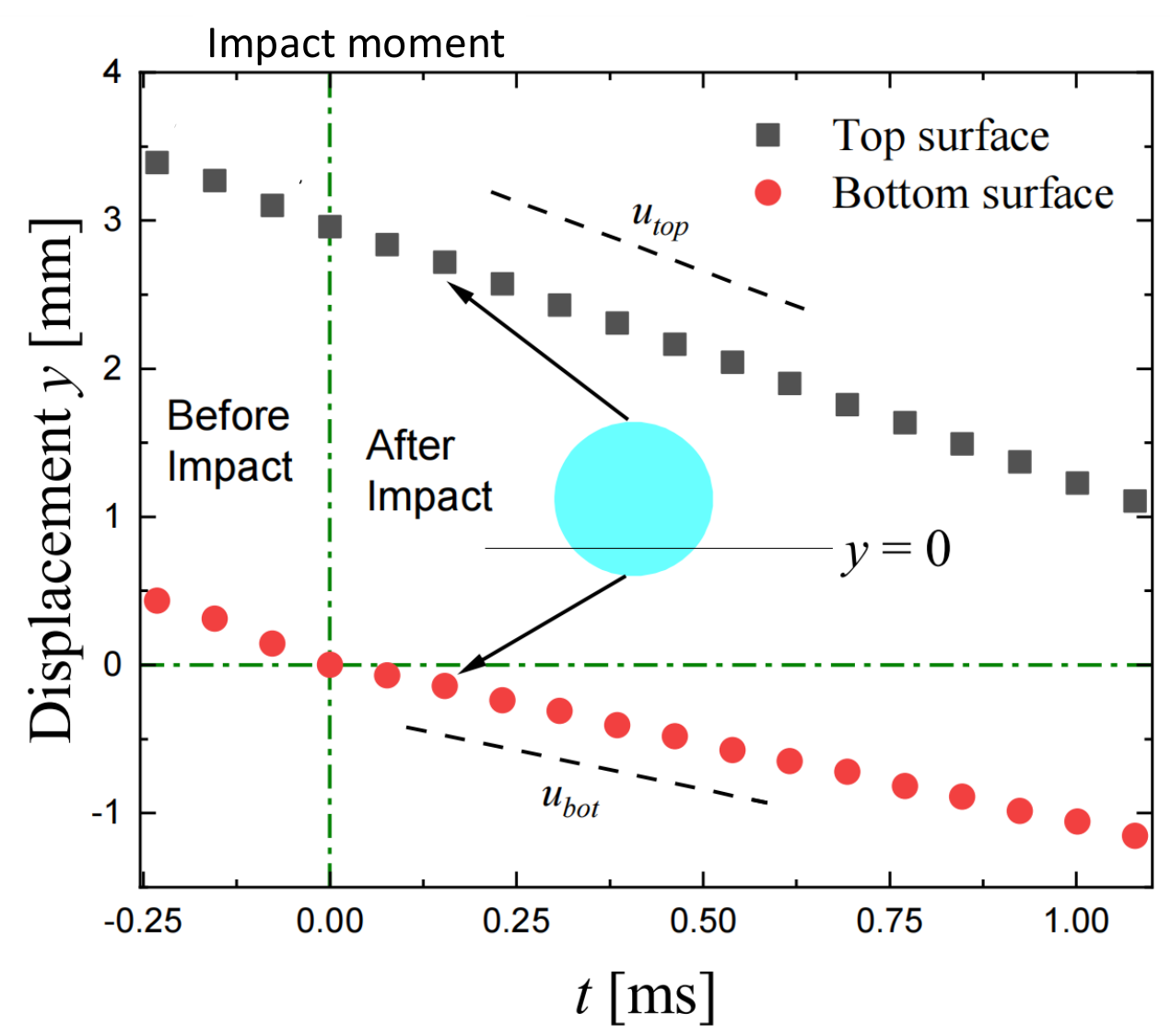}
    \caption{Representative trajectories of the top most and bottom most points of the merged droplet during penetration into the pool. Linear fits to these trajectories yield the (approximately constant) velocities $u_{top}$ and $u_{bot}$ from the corresponding slopes.} \label{fig_ubot_utop_mearsure}
\end{figure}
\FloatBarrier

It should be noted that our measurement does not introduce a lensing effect. All of our measurements are conducted outside the pool cavity which avoids the curvature between the air and the liquid pool. Specifically, the cavity essentially is the top surface of the droplet expanding into the liquid pool, represented by the red line in Fig.8(a) of the main text. The mass from the droplet is thus outside of the cavity (blue area in Fig.8(a)), exempted from the lensing effect.
In addition, the wall of the pool chamber is flat.
Therefore, there is no lensing effect for our measurements. 

\section{\label{Sup3} Quantitative distinction between \textit{Partially Attached} and \textit{Fully Attached} regimes}

\begin{figure}[htb]
	\centering
    \includegraphics[width=0.6\linewidth]{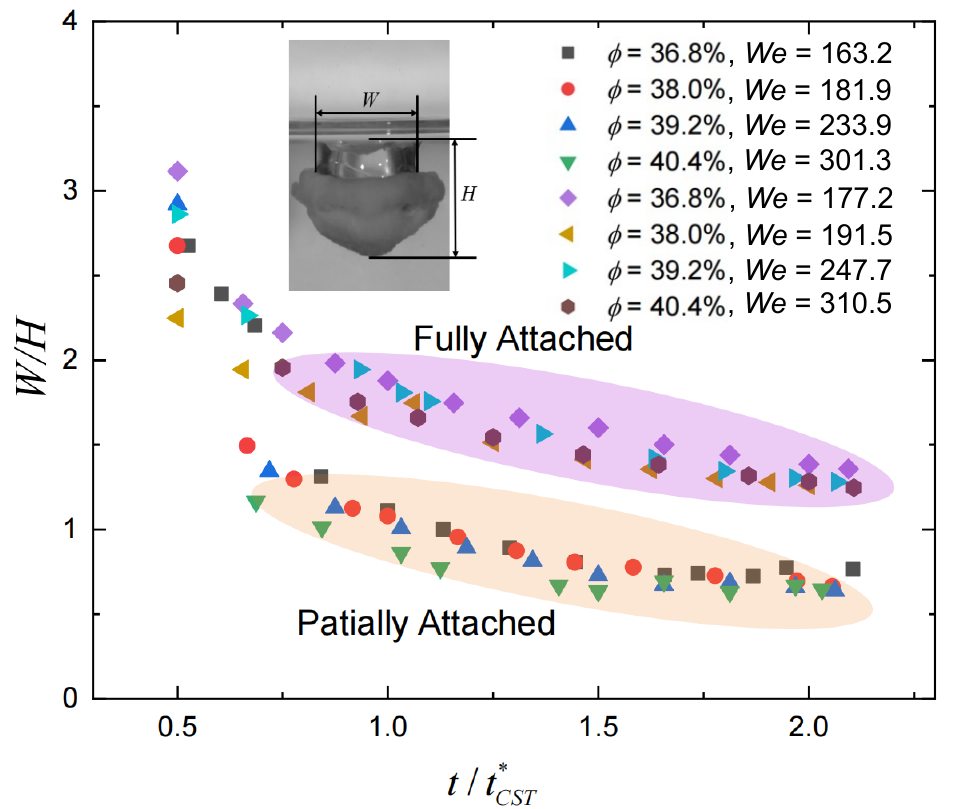}
    %{figure/W_cavity.png}
    \caption{Temporal variation of $W/H$ with normalized time $t/t_{CST}^*$, where the width $W$ and depth $H$ of the cavity are defined in the inset. The 
    ($\phi, We$) conditions are shown by  color.} \label{fig1}
\end{figure}

The quantitative distinction between \textit{Partially Attached} and \textit{Fully Attached} 
regimes is determined from the experimental results based on the evolution of the cavity 
shape. As shown in Fig.\ref{fig1}, we examine the temporal evolution of the aspect ratio $W/H$, defined as the ratio of the cavity width $W$ to the cavity height $H$, for cases in the vicinity of the transition boundary.
For the \textit{Fully Attached} regime, $W/H \approx 2$ at $t = t_{CST}^*$ (the moment at which the cavity meets the bottom of the mass from the droplet) indicating that the cavity adopts a nearly hemispherical geometry. This result is consistent with the Newtonian droplet impact scenario.
In contrast, for the \textit{Partially Attached} regime, $W/H \approx 1$ at $t = t_{CST}^*$, corresponding to a nearly cylindrical cavity morphology that departs from Newtonian behavior and cannot be accounted for by classical Newtonian droplet impact theory.
Therefore, it is clear that this ratio $W/H$ exhibits distinct trends for different cases, making it a reasonable criterion to distinguish between these two regimes.

\FloatBarrier

\section{\label{Sup6}Impact of water droplet on water pool}
\begin{figure}[htb]
	\centering
	\begin{subfigure}{1.0\linewidth}
		\centering
		\includegraphics[width=0.95\linewidth]{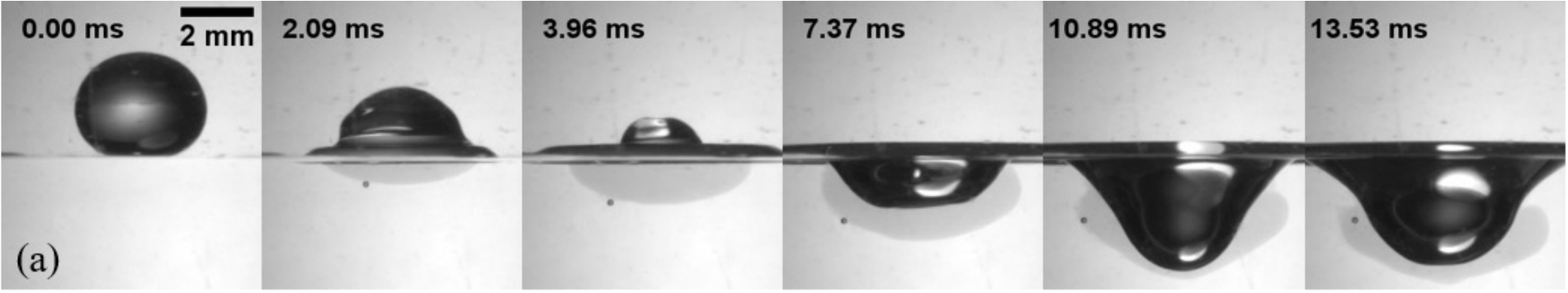}
	\end{subfigure}
	\begin{subfigure}{1.0\linewidth}
		\centering
		\includegraphics[width=0.95\linewidth]{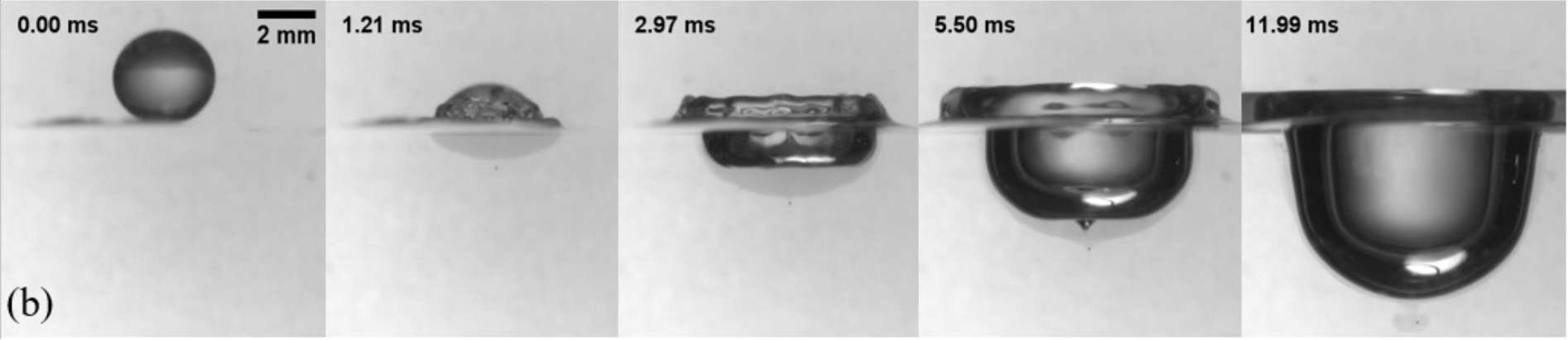}
	\end{subfigure}
    \begin{subfigure}{1.0\linewidth}
		\centering
		\includegraphics[width=0.95\linewidth]{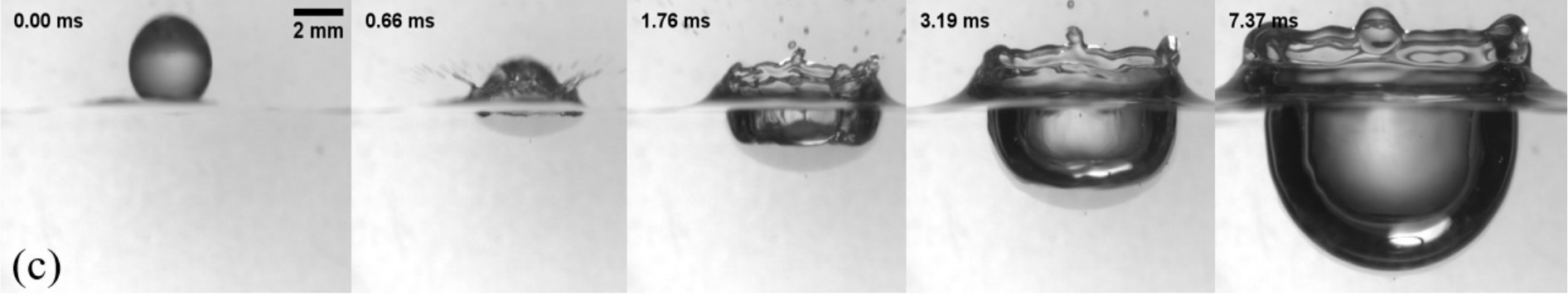}
	\end{subfigure}
    \caption{Dyed water droplets impacting a water pool at increasing Weber number: (a) $We=74$, (b) $We=157$, and (c) $We=214$. Shading tracks the mass of the dyed droplet.} \label{fig_water_impact}
\end{figure}

The impact of a water droplet on a water pool provides a Newtonian reference, and the droplet is dyed to track the mass from the droplet after merging with the pool. 
%In Fig.~\ref{fig_water_impact}(a–c), the shaded region indicates the portion of the droplet containing dye for tracking the mass from the droplet. 
At low Weber number (Fig.~\ref{fig_water_impact}(a)), the cone-shaped cavity signifies the capillary controlled dynamics and does not stretch the mass from the droplet into a thin film while subsequent cavity expansion is also absent, which is analogous to the \textit{Liquid Lump} regime for cornstarch droplets.
At higher Weber numbers (Fig.~\ref{fig_water_impact}(b,c)), the cavity deepens and expands radially to an approximately hemispherical shape, stretching the dyed mass from the droplet into a continuous thin layer, analogous to the \textit{Partially Attached} and \textit{Fully Attached} regimes observed for cornstarch droplets.

\FloatBarrier

\section{\label{Sup5}Impact dynamics of glycerol droplet on water pool}

\begin{figure}[htb]
	\centering
	\begin{subfigure}{1.0\linewidth}
		\centering
		\includegraphics[width=0.95\linewidth]{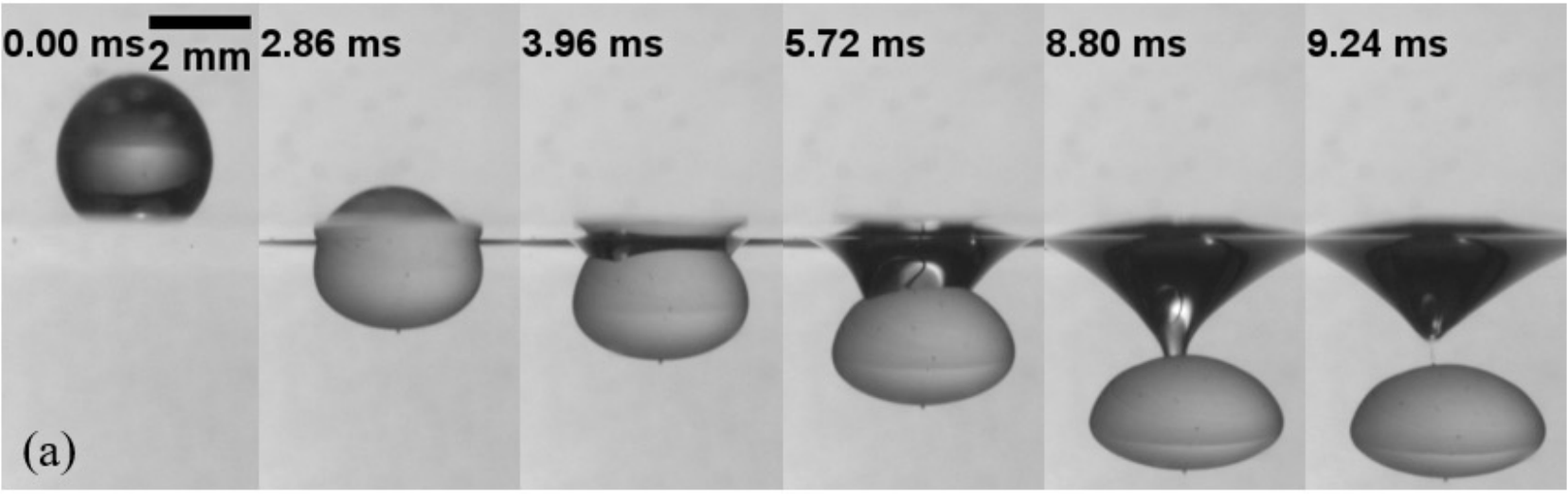}
	\end{subfigure}
    \begin{subfigure}{1.0\linewidth}
		\centering
		\includegraphics[width=0.95\linewidth]{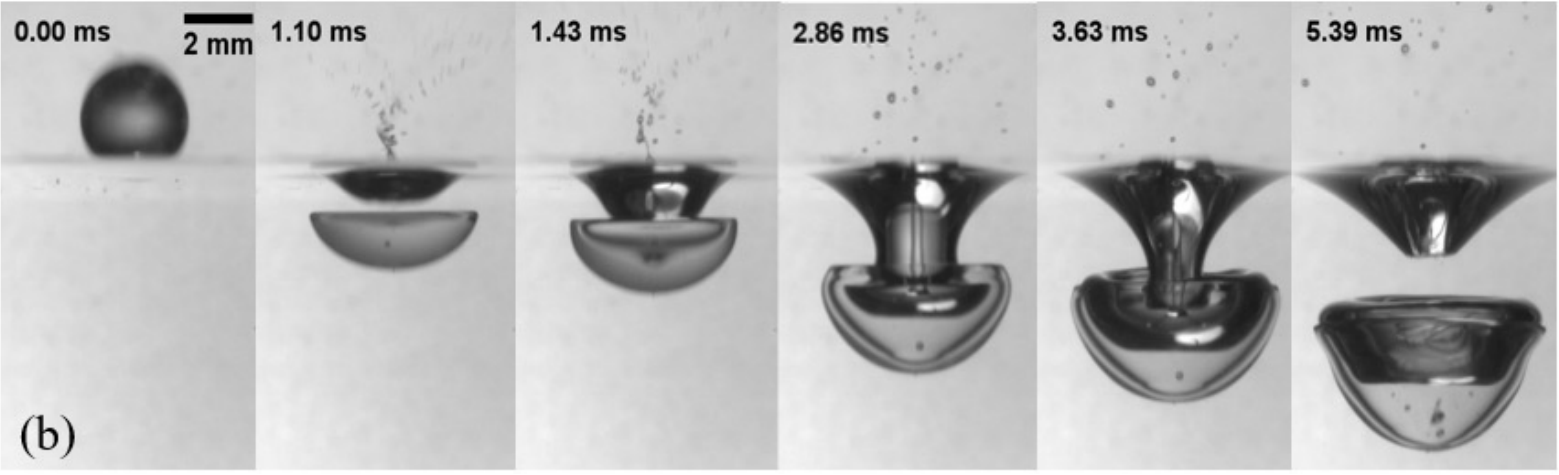}
	\end{subfigure}
    \caption{Glycerol-water droplets impacting a water pool: (a) $We=22.11$ and (b) $We=912.03$. The higher-$We$ case shows droplet--pool separation followed by the formation of a single surface-adjacent bubble.} \label{fig_glycerol_impact3}
\end{figure}

Fig.~\ref{fig_glycerol_impact3} shows the impact dynamics of a glycerol-water droplet on a water pool with a viscosity of 1000 mPa$\cdot$s, matching that in the BST region. The comparison of the impact dynamics with that of cornstarch suspension droplets highlights that the \textit{Wrapped Bubble} regime is not solely due to high viscosity and that even when a bubble forms for the glycerol-water droplet, the formation mechanism is different from that of suspension droplets. For the impact of glycerol-water droplet, bubbles do not form until much higher $We$ than for the wrapped bubble regime for the cornstarch droplet. In the glycerol-water case, once the droplet and pool surfaces separate, a single bubble forms near the pool surface (Fig.~\ref{fig_glycerol_impact3}(b), $t=5.39$ ms).
Moreover, a glycerol droplet spreads into a thin film and drives a deep cavity. In contrast, for cornstarch droplet impacts, impact-induced jamming near the bottom of the droplet suppresses cavity expansion, and the upper liquid region wraps bubbles around next to the jammed mass. Consequently, the \textit{Wrapped Bubble} regime occurs at Weber numbers much lower for cornstarch droplets than those for glycerol.

\FloatBarrier

\section{\label{SupKphi} Calculation of $k_{CST}$ and $k_{BST}$}

\begin{figure}[htb]
	\centering
	\includegraphics[width=0.95\linewidth]{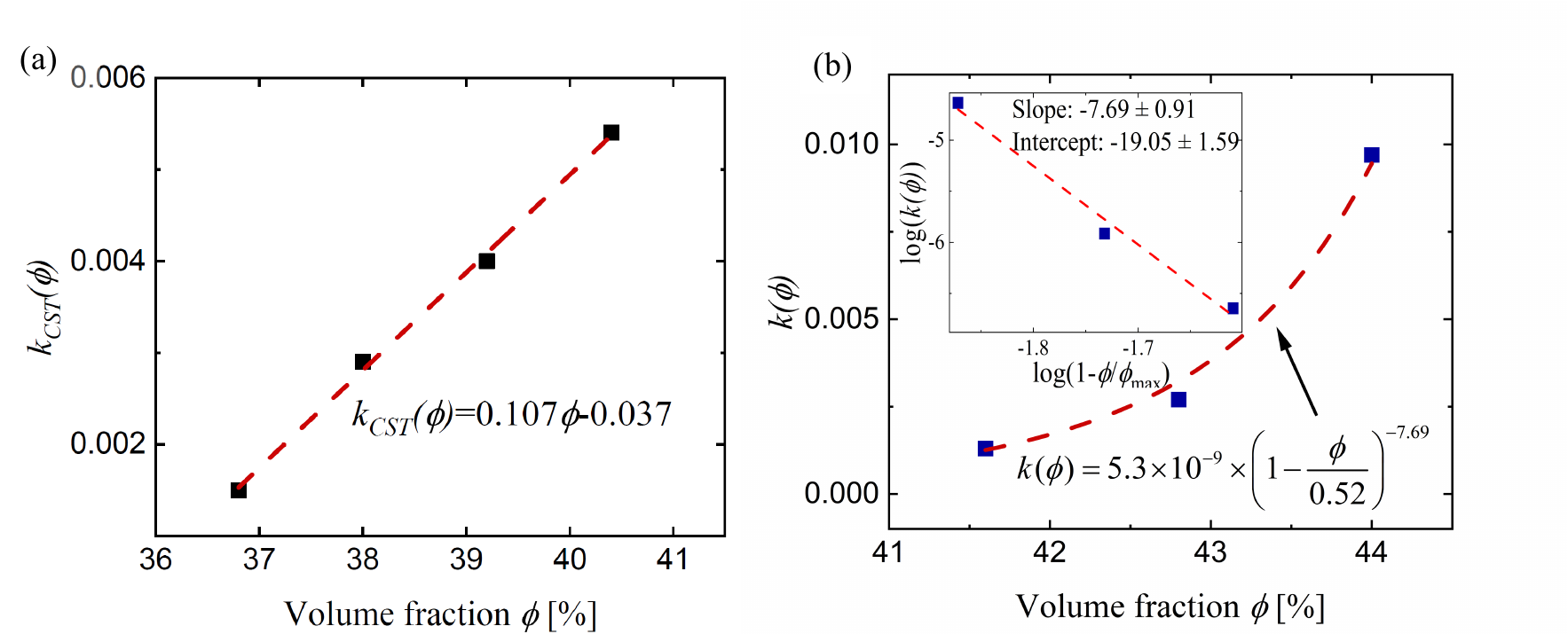}
    \caption{The data and fitting expressions for (a) $k_{CST}(\phi)$, and (b) $k_{BST}(\phi)$} \label{fig11}
\end{figure}

At low volume fraction in the CST region, the shear stress is modeled as: $\tau_\eta = k_{CST}(\phi) \dot{\gamma}^{1.5}$, where $k_{CST}(\phi)$ is extracted from the rheology data presented in Fig.6(a) of the main text.
The values of $k_{CST}(\phi)$ are plotted in Fig.~\ref{fig11}(a) and are fitted linearly: $k_{CST}(\phi )=a\phi -b$ with best-fit parameters \(a = 0.107 \pm 0.003\) $\mathrm{Pa}\cdot \mathrm{s}^{1.5}$ and \(b = 0.037 \pm 0.001\) $\mathrm{Pa}\cdot \mathrm{s}^{1.5}$. 

As for intermediate volume fractions in the BST region, $k_{BST}(\phi)$ is extracted from the rheology data based on ${{\tau }_{\eta }} = k_{BST}(\phi)\dot{\gamma}^{2}$. 
As illustrated in Fig. \ref{fig11}(b), $k_{BST}(\phi)$ cannot be approximated by a linear function as in the CST region. Instead, a classical model, $k_{BST}(\phi )=K{{\left( 1-{\phi }/{{{\phi }_{\max }}} \right)}^{\beta }}$ \cite{mewis2012colloidal}, is adopted, where $\phi_{max} = 52.0\%$ as tested, and $K, \beta$ are fitting parameters.
$k_{BST}(\phi)$ is fitted as $\text{log}(k_{BST}(\phi)) = \text{log}K + \beta\text{log}(1-\phi/{{\phi }_{\max }})$ from rheology data shown in Fig.6(a) to give $\beta = -7.69 \pm 0.91$ and the numerical value of the $\text{log}K$ is $-19.06 \pm 1.59$. 
Thus, $k_{BST}(\phi) = 5.3\times {{10}^{-9}}(1-\phi/0.52)^{-7.69}$ $\mathrm{Pa}\cdot \mathrm{s}^{2}$ is used in the energy analysis in the BST region.

\bibliographystyle{unsrtnat}
\bibliography{apssamp_f}